\pdfoutput=1

\documentclass[11pt]{article}

\usepackage[table,xcdraw]{xcolor}

\usepackage[]{acl}

\usepackage{times}
\usepackage{latexsym}

\usepackage[T1]{fontenc}

\usepackage[utf8]{inputenc}
\usepackage{enumitem}
\setlist[itemize]{noitemsep, topsep=0pt, leftmargin=*}

\usepackage{microtype}

\usepackage{inconsolata}

\usepackage{microtype}
\usepackage{hyperref}
\usepackage{url}
\usepackage{booktabs}
\usepackage{lipsum}
\usepackage{booktabs}
\usepackage{multirow}
\usepackage{graphicx}
\usepackage{tabularx}
\usepackage{textcomp}
\usepackage{xspace}
\usepackage{pifont}
\usepackage{amssymb} 
\usepackage[stable]{footmisc} 
\newcommand{\model}{\textit{OpenCodeInterpreter}\xspace}
\newcommand{\dataset}{Code-Feedback\xspace}
\usepackage{arydshln} 
\usepackage{listings} 
\usepackage{xcolor} 
\usepackage{hyperref}
\usepackage{caption}
\usepackage{tcolorbox}
\usepackage{soul}
\usepackage{mdframed}
\usepackage{listings}
\usepackage{courier}
\usepackage{tcolorbox}

\definecolor{softblue}{rgb}{0.88, 0.95, 1.0} 
\definecolor{softyellow}{rgb}{0.98, 0.98, 0.82} 

\newcommand{\highlightyellow}[1]{\sethlcolor{softyellow}\hl{#1}}
\newcommand{\highlightblue}[1]{\sethlcolor{softblue}\hl{#1}}

\newcommand*\colourcheck[1]{%
  \expandafter\newcommand\csname #1check\endcsname{\textcolor{#1}{\ding{52}}}%
}
\colourcheck{blue}
\colourcheck{green}
\colourcheck{red}

\newcommand*\colourmark[1]{%
  \expandafter\newcommand\csname #1mark\endcsname{\textcolor{#1}{\ding{55}}}%
}
\colourmark{blue}
\colourmark{green}
\colourmark{red}

\title{\model: Integrating Code Generation with \\ Execution and Refinement}

\author{Tianyu Zheng$^1$\thanks{Equal Contributions.}, Ge Zhang$^{1,2}$\footnotemark[1], Tianhao Shen$^{1}$\footnotemark[1], Xueling Liu$^1$\footnotemark[1],\\
    \textbf{Bill Yuchen Lin$^3$, Jie Fu$^{4}$, Wenhu Chen$^{1,2}$, Xiang Yue$^{1,5}$\thanks{Corresponding Author.}}
    \\[3pt]
    $^1$Multimodal Art Projection Research Community, $^2$University of Waterloo,\\ $^3$Allen Institute for Artificial Intelligence, $^4$HKUST,
    $^{5}$IN.AI Research\\[1pt]
\small \texttt{\{zhengtianyu0428, xiangyue.work\}@gmail.com, ge.zhang@uwaterloo.ca} \\[2pt]
\vspace{-4ex}
\url{https://opencodeinterpreter.github.io}
}

\begin{document}
\maketitle
\begin{abstract}
The introduction of large language models has significantly advanced code generation. However, open-source models often lack the execution capabilities and iterative refinement of advanced systems like the GPT-4 Code Interpreter. To address this, we introduce \model, a family of open-source code systems designed for generating, executing, and iteratively refining code. Supported by \dataset, a dataset featuring 68K multi-turn interactions, \model integrates execution and human feedback for dynamic code refinement. Our comprehensive evaluation of \model across key benchmarks such as HumanEval, MBPP, and their enhanced versions from EvalPlus reveals its exceptional performance. Notably, \model-33B achieves an accuracy of 83.2 (76.4) on the average (and plus versions) of HumanEval and MBPP, closely rivaling GPT-4's 84.2 (76.2) and further elevates to 91.6 (84.6) with synthesized human feedback from GPT-4. \model brings the gap between open-source code generation models and proprietary systems like GPT-4 Code Interpreter.

\end{abstract}

\begin{figure}[!t]
    \centering
\includegraphics[width=\linewidth]{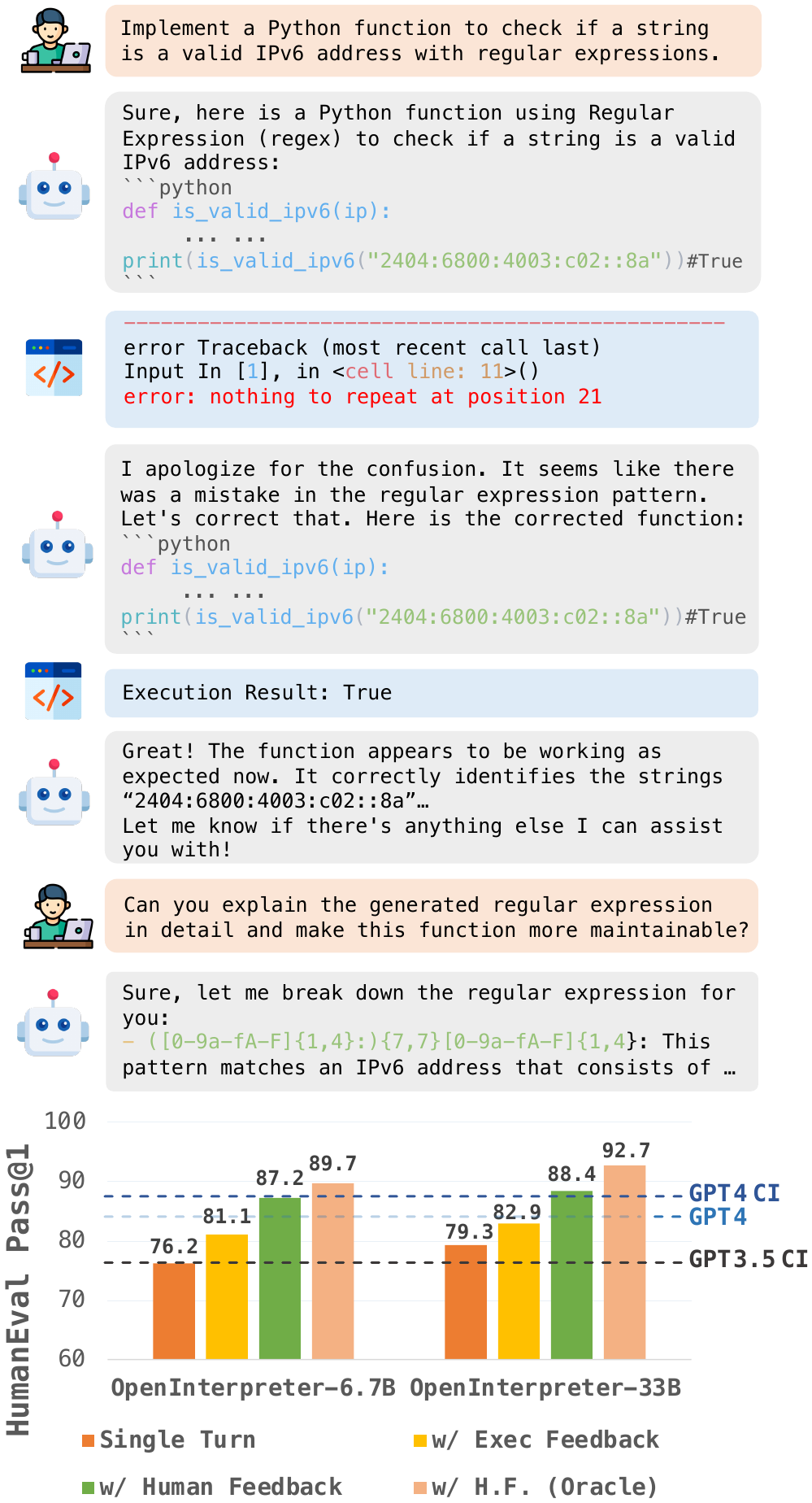}
    \caption{Overview of the \model and its pass@1 accuracy on the HumanEval. With appropriate feedback, \model-33B achieves performance comparable to that of the GPT-4 Code Interpreter. }
    \label{fig:teaser}
    \vspace{-2ex}
\end{figure}
\section{Introduction}

Code generation has been a pivotal challenge within computer science for several decades. 
Recently, the landscape of code generation has been revolutionized by the advent of large language models (LLMs) pre-trained on extensive code corpora~\citep{nijkamp2022codegen,christopoulou2022pangu,zheng2023codegeex,li2023starcoder,wang2023codet5+,roziere2023code,guo2024deepseek}. 
These models have showcased remarkable capabilities in generating code that accurately aligns with user intents, thus providing substantial support for software development~\citep{GitHubCopilot2023}. 

To unleash the capabilities of pre-trained code models, instruction-tuning methods have been developed. For instance, CodeAlpaca~\citep{chaudhary2023codealpaca} comprises 20K code instructions automatically generated by applying self-instruct~\citep{wang2023self} to ChatGPT, utilizing 21 seed tasks as the foundation. To further refine the coding proficiency of LLMs, \citet{luo2023wizardcoder} introduces Code Evol-Instruct, a method that applies a variety of heuristics to enrich the complexity of initial code instructions, building upon the dataset provided by CodeAlpaca. Meanwhile, MagicCoder~\citep{wei2023magicoder} employs a robust LLM to generate novel coding challenges, sourcing inspiration from a diverse range of open-source code snippets. Additionally, WaveCoder~\citep{yu2023wavecoder} implements an LLM generator-discriminator framework for creating code instruction data, offering customization and control over the data generation process.

Despite these advancements, current code models are constrained by their capacity to utilize feedback for refinement. Essentially, feedback can have two forms: (1) \textit{execution feedback}, which includes execution outputs and diagnostics, and (2) \textit{human feedback}, comprising follow-up guidance or instructions from users. Execution feedback plays a vital role in enabling models to rectify syntactic and logical errors, and human feedback aids models in better understanding user instructions, facilitating the generation of solutions that more closely align with user expectations.

To address these challenges, we propose \model, a family of open-source code systems designed for generating, executing, and iteratively refining code. 
\model is trained on our constructed \dataset dataset, which features 68K multi-turn interactions between users, code models, and compilers. \model uniquely integrates both execution and human feedback, employing compiler diagnostics to rectify errors and human insights to refine code generation. This approach allows \model to produce solutions that are both technically sound and closely matched to user requirements, significantly boosting its overall performance.

Our thorough evaluation of \model on widely recognized benchmarks, such as HumanEval~\cite{chen2021evaluating}, MBPP~\cite{austin2021program}, and their augmented counterparts from EvalPlus~\cite{liu2023your}, highlights its superior ability to generate and iteratively refine code, achieving exemplary standards of quality and functionality. Remarkably, \model-33B secures an impressive accuracy of 83.2 (76.4) on the average (and plus versions) of HumanEval and MBPP, showcasing performance on par with GPT-4's 84.2 (76.2). Furthermore, when augmented with synthesized human feedback from GPT-4, \model's performance notably increases to 91.6 (84.6). \model thereby establishes a new benchmark in code generation, effectively narrowing the performance gap between open-source models and sophisticated proprietary systems like the GPT-4 Code Interpreter.

\section{\dataset}

\begin{figure*}[!t]
    \centering 
    \begin{minipage}{0.40\linewidth}
        \centering
        \includegraphics[width=\linewidth]{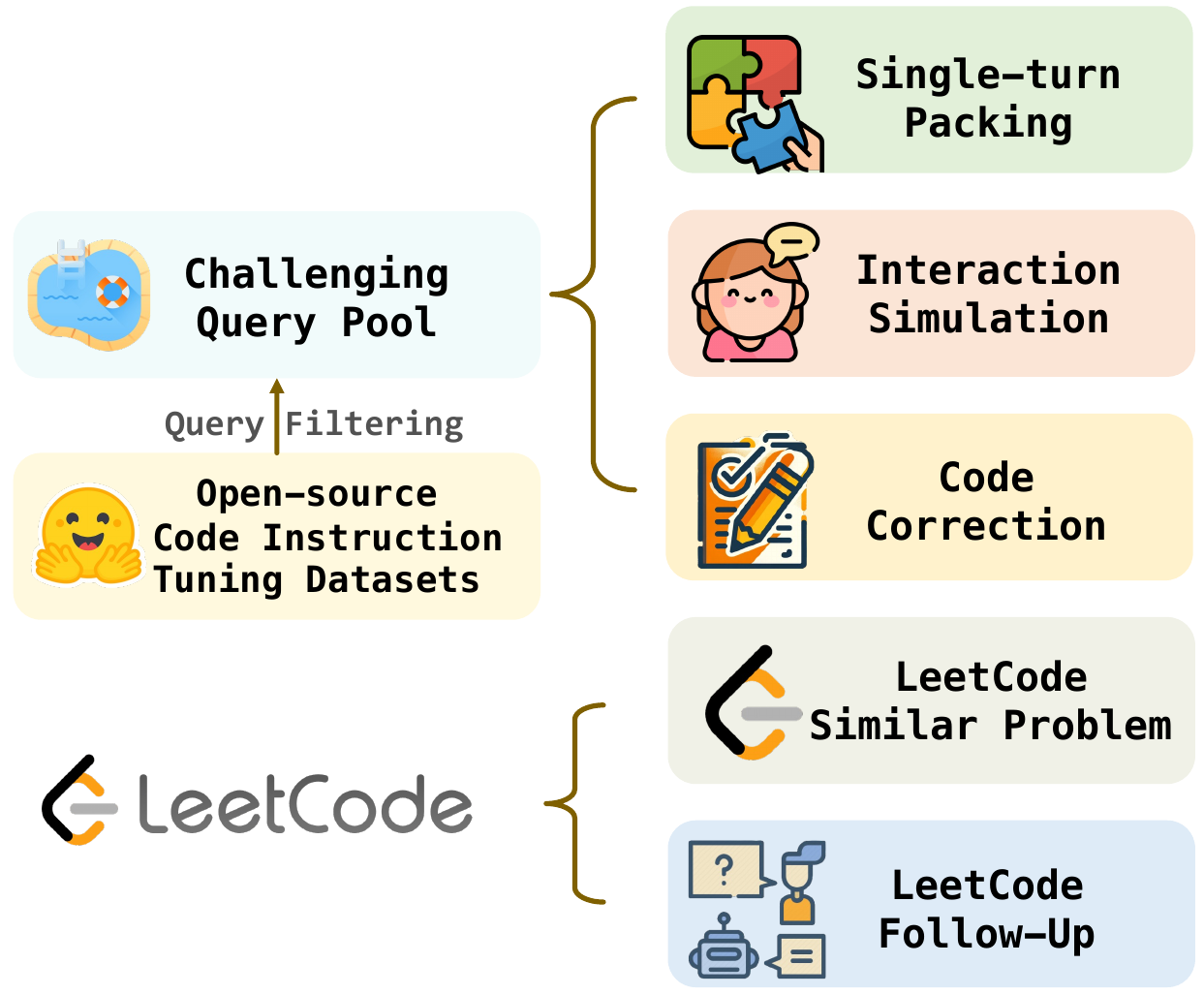} 
    \end{minipage}%
    \hfill
    \begin{minipage}{0.56\linewidth}
        \resizebox{\linewidth}{!}{%
        \begin{tabular}{@{}lcccccc@{}}
        \toprule
        Dataset             & \#Sample & \#Turn & M.T & E.F & H.F  \\
        \midrule
        CodeAlpaca\footref{fn:alpacacodes}  & 20k    & 20K & \redmark & \redmark & \redmark \\
         Magicoder-OSS-Instruct\footref{magicoder}  & 75K     & 75K & \redmark & \redmark & \redmark \\
         Python-Code-ShareGPT\footref{sharegpt}   & 23K     & 23K & \redmark & \redmark & \redmark \\
         Magicoder-Evol-Instruct\footref{magicoderevol}  & 111K    & 111K & \redmark & \redmark & \redmark \\
         EvolInstruct-Code\footref{evolinstruct}      & 80k    & 80K & \redmark & \redmark & \redmark \\\midrule
         \textbf{Code-Feedback (Ours)}     & 68K     & 192K & \greencheck & \greencheck & \greencheck \\[2pt]\hdashline\\[-8pt]
         \quad Single-turn Packing      & 16K     & 33.5K & \greencheck & \redmark & \greencheck\\
         \quad Interaction Simulation   & 51K     & 155.5K & \greencheck & \greencheck & \greencheck \\
         \quad Code Correction   & 0.5K    & 1.2K   & \greencheck & \greencheck & \redmark \\
         \quad LeetCode Similar Problem      & 0.3K    & 0.65K & \greencheck & \redmark & \greencheck\\
         \quad LeetCode Follow-Up    & 0.2K    & 0.76K & \greencheck & \redmark & \greencheck \\
        \bottomrule
        \end{tabular}%
        }
        
    \end{minipage}
            \caption{Summary of our proposed dataset \dataset construction and comparison with existing code instruction tuning datasets. M.T: Multi Turn, E.F: Execute Feedback, H.F: Human Feedback.}
\label{fig:combined_data_collection_statistics}

\end{figure*}

In this section, we detail the creation of our code instruction tuning dataset, \dataset (\autoref{fig:combined_data_collection_statistics}), designed to train \model. 
\dataset is crafted to meet specific criteria: \textbf{1) Diverse and challenging real-world queries:} The dataset should encompass a wide range of queries derived from real-world coding tasks, presenting both diversity and complexity.
\textbf{2) Multi-turn dialogue structure:} \dataset is structured as multi-turn dialogues, incorporating two types of feedback: execution feedback, which includes outputs and diagnostics from compilers, and human feedback, consisting of additional guidance or instructions from users.
\textbf{3) Interleaved text and code responses:} Each response is expected to provide responses that blend natural language explanations with code snippets, offering a holistic approach to solving coding queries.

To assemble a dataset that fulfills these desiderata, we have employed five distinct methods. 
Examples of these five categories can be found in Appendix \ref{sec:data_collection_examples}.
The sources of our queries fall into two main categories: \textit{a variety of open-source datasets} and \textit{coding challenges from LeetCode}. In the next subsections, we will discuss how we develop data construction methods to meet the three aforementioned criteria from the two data sources.

\newcommand{\alpacacodefootnote}{\footnote{\url{https://huggingface.co/datasets/HuggingFaceH4/CodeAlpaca_20K}}\label{fn:alpacacode}}
\footnotetext{\href{https://huggingface.co/datasets/HuggingFaceH4/CodeAlpaca_20K}{hf.co/datasets/HuggingFaceH4/CodeAlpaca\_20K}\label{fn:alpacacodes}}

\subsection{Coding Queries from Open-source Data}
\label{sec:coding_queries_from_open_source_data}
We have aggregated 287k queries from four distinguished open-source code instruction tuning datasets: Magicoder-OSS-Instruct\footnote{\href{https://huggingface.co/datasets/ise-uiuc/Magicoder-OSS-Instruct-75K}{hf.co/datasets/ise-uiuc/Magicoder-OSS-Instruct-75K}\label{magicoder}}, Python code subset of ShareGPT\footnote{\href{https://huggingface.co/datasets/ajibawa-2023/Python-Code-23k-ShareGPT}{hf.co/datasets/ajibawa-2023/Python-Code-23k-ShareGPT}\label{sharegpt}}, Magicoder-Evol-Instruct\footnote{\href{https://huggingface.co/datasets/ise-uiuc/Magicoder-Evol-Instruct-110K}{hf.co/datasets/ise-uiuc/Magicoder-Evol-Instruct-110K}\label{magicoderevol}}, and Evol-Instruct-Code\footnote{\href{https://huggingface.co/datasets/nickrosh/Evol-Instruct-Code-80k-v1}{hf.co/datasets/nickrosh/Evol-Instruct-Code-80k-v1}\label{evolinstruct}}. To refine this extensive collection and isolate the most intricate and informative instructions, we employ a very capable open-source chat model, Qwen-72B-Chat~\cite{bai2023qwen}, for a selective filtering process. 
This involves the LLM assessing each code query and its corresponding response within the compiled datasets on a complexity score from 1 to 5. 
Only the most challenging queries, with ratings of 4 or 5, were retained for our seed set, ensuring a focus on the most difficult instructions.
To guarantee the robustness of our selection, this filtering operation is repeated with two distinct prompts (detailed in Appendix~\ref{sec:prompt_used_in_data_filtering}), thereby solidifying the complexity of our final query selection. This meticulous process resulted in 156k high-quality single-turn code instructions as the challenging query pool. Detailed statistics of this data compilation are provided in {Appendix~\ref{sec:prompt_used_in_data_filtering}}.

Subsequently, we describe three methods employed to transform this curated single-turn data into multi-turn dialogues enriched with both execution and human feedback.

\noindent\textbf{Singe-turn Packing.} A direct approach to crafting multi-turn data is to group single-turn query-response pairs into multi-turn formats. Inspired by in-context pre-training techniques~\cite{shi2023context}, which consolidate similar sequences to foster model learning of dependencies among related documents, we merge similar single-turn query-response pairs to form multi-turn dialogues.

Utilizing the BERT-base embedding~\cite{devlin2018bert}, we convert queries into vectorized representations. For each query, the $k$-nearest neighbors algorithm is employed to identify its four closest counterparts. From these, we randomly select two or three to assemble multi-turn sequences. To maintain data uniqueness, once a query is chosen as a neighbor, it is exempt from future selections as a neighboring query, ensuring no single instruction is repeated across the dataset. Should a query's potential neighbors have been previously utilized, that query is bypassed. This method results in the creation of 16.6K multi-turn instances derived from 105K single-turn instances.

\noindent\textbf{Interaction Simulation.} Gathering authentic human interaction data poses significant challenges. To replicate a realistic code interpreter usage scenario, we developed a simulator using GPT-3.5 and GPT-4. For each selected query, GPT-3.5 first generates a preliminary response from which we extract the code snippet and execute it. The outcome of this execution, along with any compiler diagnostics, is then fed into GPT-4 to elicit a follow-up response. This cycle is repeated until GPT-4 delivers what it deems a correct solution or until a maximum of three iterations is reached.

Subsequently, we introduce simulated human feedback into the interaction. We predefine ten common feedback categories, including issues related to syntax and formatting, efficiency, functionality, clarity, bugs, security, compatibility, resource use, scalability, and best practices, with each category detailed in Appendix \ref{sec:prompt_used_in_data_collection}. GPT-4 is then prompted to select the most relevant feedback for the scenario and generate appropriate responses within that feedback category. By incorporating this simulated feedback into the dialogue history, GPT-4 is encouraged to refine its solutions further, mimicking intricate user-model exchanges and demonstrating self-correction in response to human input. Through this simulation approach, we have constructed 51K examples, effectively capturing the nuanced dynamics of user interactions and feedback-driven solution refinement.

\noindent\textbf{Code Correction.}
To boost the model's error-handling capabilities, we include a focused stage in our data compilation that generates 500 specific error correction interactions. We initiate this by prompting GPT-4 to \textit{intentionally} produce incorrect code snippets, as outlined in Appendix \ref{sec:prompt_used_in_data_collection}. The model then uses the error messages from executing these snippets as cues for corrections. This approach mirrors the real-life coding cycle, where developers continuously debug and refine their code, thus enriching our dataset with a broad spectrum of error correction examples. Following this, we replace the initial prompts that resulted in incorrect code with the ones that encourage the generation of correct code outputs. This method ensures the model learns from both successful code generation and error identification and correction, significantly enhancing its problem-solving skills and understanding of the debugging process.

\subsection{Coding Challenges from LeetCode}
\label{sec:coding_challenges_from_leetcode}
\noindent\textbf{LeetCode Similar Problem.}
Drawing inspiration from the practice among programmers of honing their skills through LeetCode challenges, we gather similar LeetCode questions and their solutions from the TACO dataset~\cite{li2023taco}. LeetCode\footnote{\url{https://leetcode.com/problemset/}} categorizes related questions through tags, facilitating the extraction of connected problems. TACO ensures the LeetCode dataset is cleansed to prevent any unintended impact on other task datasets, such as HumanEval and MBPP. By amalgamating associated LeetCode questions, we compile 303 multi-turn instances, enriching the dataset with varied coding challenges.

\noindent\textbf{LeetCode Follow-up Question.}
We further delve into the LeetCode dataset to isolate solutions to identical questions that differ in time or space complexity or are implemented in various programming languages. This process of aggregating diverse solutions to the same LeetCode questions yields 200 multi-round instances, showcasing alternative problem-solving approaches.

Given the original LeetCode solutions often lack comprehensive natural language explanations, we engage GPT-4 to enhance these solutions with integrated text explanations and code snippets, standardizing all instances into a consistent format. The specific prompts used to guide GPT-4 in this enrichment process are detailed in Appendix~\ref{sec:prompt_for_nl_explanation}, ensuring clarity and educational value in the responses.

\begin{table*}[!tbp]
\small
\centering
\scalebox{1.0}{
\begin{tabular}{@{}lllccccc@{}}
\toprule
\multirow{2}{*}{Model} & \multirow{2}{*}{Size} & \multirow{2}{*}{Type} & \multicolumn{2}{c}{Open-source} & \multirow{2}{*}{HumanEval (+)} & \multirow{2}{*}{MBPP (+)} & \multirow{2}{*}{Average (+)} \\
 &  &  & Model & Data &  &  &  \\ \midrule
 \rowcolor{softyellow}
GPT-4 Turbo &  &  &  &  & 85.4 (81.7) & 83.0 (70.7) & 84.2 (76.2) \\
 \rowcolor{softyellow}
\quad + Execution Feedback  & \multirow{-2}{*}{-} & \multirow{-2}{*}{-} & \multirow{-2}{*}{\textopenbullet} & \multirow{-2}{*}{\textopenbullet} & \textbf{88.0} (\textbf{84.2})  & \textbf{92.0} (\textbf{78.2}) & \textbf{90.0} (\textbf{81.2}) \\ 
 \rowcolor{softyellow}
GPT-3.5 Turbo &  &  &  &  & 72.6 (65.9) & 81.7 (69.4) & 77.2 (67.7) \\
 \rowcolor{softyellow}

\quad + Execution Feedback  & \multirow{-2}{*}{-} & \multirow{-2}{*}{-} & \multirow{-2}{*}{\textopenbullet} & \multirow{-2}{*}{\textopenbullet} & 76.8 (70.7) & 87.0 (73.9) & 81.9 (72.3) \\ [2pt]\hdashline\\[-8pt]

Gemini Pro~\cite{team2023gemini} & - & - & \textopenbullet & \textopenbullet & 63.4 (55.5) & 72.9 (57.9) & 68.2 (56.7) \\ \midrule
\multicolumn{8}{c}{$\sim$7B Scale} \\ \midrule
CodeT5+~\cite{wang2023codet5+} & 6B & Base & \textbullet & \textbullet & 29.3 (23.8) & 51.9 (40.9) & 40.6 (32.4) \\
CodeGen-Mono~\cite{nijkamp2022codegen} & 6B & Base & \textbullet & \textbullet & 29.3 (25.6) & 49.9 (42.1) & 39.6 (33.9) \\
OpenChat~\cite{wang2023openchat} & 7B & Instruct & \textbullet & \textbullet & 72.0 (67.1) & 62.7 (52.9) & 67.4 (60.0) \\[2pt]\hdashline\\[-8pt]

StarCoder2~\cite{lozhkov2024starcoder} & 7B & Base & \textbullet & \textbullet & 35.4 (29.9) & 54.4 (45.6) & 44.9 (37.8) \\
\rowcolor{softblue}
\quad \textbf{\model-SC2} & & & & & 73.8 (68.9)	 & 61.7 (51.1) & 67.8 (60.0)\\
\rowcolor{softblue}
\quad + Execution Feedback & \multirow{-2}{*}{7B} & \multirow{-2}{*}{Instruct} & \multirow{-2}{*}{\textbullet} & \multirow{-2}{*}{\textbullet} & 75.6 (69.5) &66.9 (55.4) & 71.3 (62.5) \\ 
CodeLlama-Python~\cite{roziere2023code} & 7B & Base & \textbullet & \textopenbullet & 37.8 (34.1) & 57.6 (45.4) & 47.7 (39.8) \\
\quad WizardCoder-CL~\cite{luo2023wizardcoder} & 7B & Instruct & \textopenbullet & \textopenbullet & 48.2 (40.9) & 56.6 (47.1) & 52.4 (44.0) \\
\quad Magicoder-CL~\cite{wei2023magicoder} & 7B & Instruct & \textbullet & \textbullet & 60.4 (55.5) & 64.2 (52.6) & 62.3 (54.1) \\
\quad Magicoders-S-CL~\cite{wei2023magicoder} & 7B & Instruct & \textbullet & \textbullet & 70.7 (66.5) & 68.4 (56.6) & 69.6 (61.6) \\
\rowcolor{softblue}
\quad \textbf{\model-CL} &  &  &  & & 72.6 (67.7) & 66.4 (55.4) & 69.5 (61.6) \\ 
\rowcolor{softblue}
\quad + Execution Feedback & \multirow{-2}{*}{7B} &  \multirow{-2}{*}{Instruct}& \multirow{-2}{*}{\textbullet} & \multirow{-2}{*}{\textbullet}  & 75.6 (70.1) & 69.9 (60.7) & 72.8 (65.4) \\[2pt]\hdashline\\[-8pt]

DeepseekCoder~\cite{guo2024deepseek} & 6.7B & Base & \textbullet & \textopenbullet & 47.6 (39.6) & 70.2 (56.6) & 58.9 (48.1) \\
\rowcolor{softyellow}
\quad DeepseekCoder-Instruct &  &  &  &  & 73.8 (70.1) & 73.2 (63.4) & 73.5 (66.8) \\
\rowcolor{softyellow}
\quad + Execution Feedback & \multirow{-2}{*}{6.7B} & \multirow{-2}{*}{Instruct} & \multirow{-2}{*}{\textbullet} & \multirow{-2}{*}{\textopenbullet}  & 80.5 (75.6) & 79.9 (70.4) & 80.2 (73.0)  \\ 

\quad Magicoder-DS~\cite{wei2023magicoder} & 6.7B & Instruct & \textbullet & \textbullet & 66.5 (60.4) & 75.4 (61.9) & 71.0 (61.2) \\

\rowcolor{softyellow}
\quad Magicoder-S-DS~\cite{wei2023magicoder} &  &  &  &  & 76.8 (70.7) & 75.7 (64.4) & 76.3 (67.6) \\
\rowcolor{softyellow}
\quad + Execution Feedback & \multirow{-2}{*}{6.7B} & \multirow{-2}{*}{Instruct} & \multirow{-2}{*}{\textbullet} & \multirow{-2}{*}{\textbullet}  & 77.4 (72.0)  & 73.2 (62.4) & 75.3 (67.2)  \\

\rowcolor{softblue}
\quad \textbf{\model-DS}  & & & & & 76.2 (72.0) & 73.9 (63.7) & 75.1 (67.9) \\
\rowcolor{softblue}
\quad + Execution Feedback  & & & & & 81.1 (78.7) & 82.7 (72.4) & 81.9 (75.6) \\
\rowcolor{softblue}
\quad + Synth. Human  Feedback  & & & & & 87.2 (\textbf{86.6}) & 86.2 (74.2) & 86.7 (80.4) \\
\rowcolor{softblue}
\quad + Synth. Human  Feedback (Oracle) & \multirow{-4}{*}{6.7B} & \multirow{-4}{*}{Instruct} & \multirow{-4}{*}{\textbullet} & \multirow{-4}{*}{\textbullet}& \textbf{89.7} (\textbf{86.6}) & \textbf{87.2} (\textbf{75.2}) & \textbf{88.5} (\textbf{80.9}) \\ \midrule

\multicolumn{8}{c}{$\sim$13B Scale} \\ \midrule
CodeGen-Mono~\cite{nijkamp2022codegen} & 16B & Base & \textbullet & \textbullet & 32.9 (27.4) & 52.6 (43.6) & 42.8 (35.5) \\
CodeT5+~\cite{wang2023codet5+} & 16B & Base & \textbullet & \textopenbullet & 31.7 (26.2) & 54.6 (44.4) & 43.2 (35.3) \\[2pt]\hdashline\\[-8pt]

StarCoder2~\cite{lozhkov2024starcoder} & 15B & Base & \textbullet & \textbullet & 46.3 (37.8) & 66.2 (53.1) & 56.3 (45.5) \\
\rowcolor{softblue}
\quad \textbf{\model-SC2} & & & & & 75.6 (69.5)	 & 71.2 (61.2) & 73.4 (65.4)\\
\rowcolor{softblue}
\quad + Execution Feedback & \multirow{-2}{*}{15B} & \multirow{-2}{*}{Instruct} & \multirow{-2}{*}{\textbullet} & \multirow{-2}{*}{\textbullet} & 77.4 (72.0) &74.2 (63.4) & 75.8 (67.7) \\

CodeLlama-Python~\cite{roziere2023code} & 13B & Base & \textbullet & \textopenbullet & 42.7 (36.6) & 61.2 (50.9) & 52.0 (43.8) \\
\rowcolor{softblue}
\quad \textbf{\model-CL} & & & & & 77.4 (73.8) & 70.7 (59.2) & 74.1 (66.5) \\
\rowcolor{softblue}
\quad + Execution Feedback & \multirow{-2}{*}{13B} & \multirow{-2}{*}{Instruct} & \multirow{-2}{*}{\textbullet} & \multirow{-2}{*}{\textbullet} & \textbf{81.1} (\textbf{76.8}) & \textbf{78.2} (\textbf{67.2}) & \textbf{79.7} (\textbf{72.0}) \\ \midrule

\multicolumn{8}{c}{$\sim$34B Scale} \\ \midrule

CodeLlama-Python~\cite{roziere2023code} & 34B & Base & \textbullet & \textopenbullet & 51.8 (43.9) & 67.2 (52.9) & 59.5 (48.4) \\
\quad Speechless-CL-v2.0~\cite{speechless} & 34B & Instruct & \textbullet & \textbullet & 77.4 (71.3) & 72.4 (59.1) & 74.9 (65.2) \\
\quad XwinCoder-CL~\cite{xwin-lm} & 34B & Instruct & \textbullet & \textbullet & 75.6 (67.7) & 76.2 (62.4) & 75.9 (65.1) \\
\quad Phind-CL-v2~\cite{Phind} & 34B & Instruct & \textbullet & \textopenbullet & 71.3 (67.1) & - & - \\
\quad WizardCoder-CL~\cite{luo2023wizardcoder} & 34B & Instruct & \textbullet & \textopenbullet & 73.2 (64.6) & 73.2 (59.9) & 73.2 (62.3) \\
\rowcolor{softblue}
\quad \textbf{\model-CL} & & & & & 78.0 (72.6) & 73.4 (61.4) & 75.7 (67.0) \\
\rowcolor{softblue}
\quad + Execution Feedback & \multirow{-2}{*}{34B} & \multirow{-2}{*}{Instruct} & \multirow{-2}{*}{\textbullet} & \multirow{-2}{*}{\textbullet}& 81.7 (78.7) & 80.2 (67.9)& 81.0 (73.3)  \\\hdashline\\[-8pt]

DeepSeekCoder~\cite{guo2024deepseek} & 33B & Base & \textbullet & \textopenbullet & 51.2 (44.5) & - & - \\
\rowcolor{softyellow}
\quad DeepSeekCoder-Instruct &  &  &  &  & 81.1 (75.0) & 78.7 (66.7) & 79.9 (70.9) \\
\rowcolor{softyellow}
\quad + Execution Feedback & \multirow{-2}{*}{33B} & \multirow{-2}{*}{Instruct} & \multirow{-2}{*}{\textbullet} & \multirow{-2}{*}{\textopenbullet}  & 81.1 (76.2) & 82.7 (73.4) & 81.9 (74.8)  \\ 


\rowcolor{softyellow}
\quad WizardCoder-V1.1~\cite{luo2023wizardcoder} &  &  &  &  & 79.9 (73.2) & 78.9 (66.9) & 79.4 (70.1) \\
\rowcolor{softyellow}
\quad + Execution Feedback & \multirow{-2}{*}{33B} & \multirow{-2}{*}{Instruct} & \multirow{-2}{*}{\textbullet} & \multirow{-2}{*}{\textopenbullet}  & 74.4 (69.5)  & 79.9 (68.2) & 77.2 (68.9)  \\ 

\rowcolor{softblue}
\quad \textbf{\model-DS} & & & & & 79.3 (74.3) & 78.7 (66.4) & 79.0 (70.4) \\
\rowcolor{softblue}
\quad + Execution Feedback & & & & & 82.9 (80.5) & 83.5 (72.2) & 83.2 (76.4) \\
\rowcolor{softblue}
\quad + Synth. Human  Feedback  & & & & & 88.4 (86.0) & 87.5 (75.9) & 88.0 (81.0) \\
\rowcolor{softblue}
\quad + Synth. Human  Feedback (Oracle) & \multirow{-4}{*}{33B} & \multirow{-4}{*}{Instruct} & \multirow{-4}{*}{\textbullet} & \multirow{-4}{*}{\textbullet} & \textbf{92.7} (\textbf{89.7}) & \textbf{90.5} (\textbf{79.5}) & \textbf{91.6} (\textbf{84.6}) \\ \midrule

\multicolumn{8}{c}{$\sim$70B Scale} \\ \midrule
CodeLlama-Python~\cite{roziere2023code} & 70B & Base & \textbullet & \textopenbullet & 55.5 (50.0) & 65.4 (53.4) & 60.5 (51.7) \\
\quad CodeLlama-Instruct & 70B & Instruct & \textbullet & \textopenbullet & 72.0 (65.2) & 75.4 (61.7) & 73.7 (63.5) \\
\rowcolor{softblue}
\quad \textbf{\model-CL}& & & & & 76.2 (70.7) & 73.0 (61.9) & 74.6 (66.3) \\
\rowcolor{softblue}
\quad + Execution Feedback & \multirow{-2}{*}{70B} & \multirow{-2}{*}{Instruct} & \multirow{-2}{*}{\textbullet} & \multirow{-2}{*}{\textbullet}  & \textbf{79.9} (\textbf{77.4}) & \textbf{81.5} (\textbf{69.9}) & \textbf{80.7} (\textbf{73.7}) \\ \bottomrule
\end{tabular}}%
\caption{Pass@1 accuracy of different code models on HumanEval (+), MBPP (+) and their average (+). 
`CL': based on CodeLlama; `DS': based on DeepseekCoder. 
Baseline results are copied from the EvalPlus Leaderboard or replicated by running the official checkpoints. We highlight \highlightyellow{strong baselines} and \highlightblue{our methods} for each scale. }
\label{tab:overall_results}
\end{table*}

\section{Experimental Setup}
\noindent\textbf{Training Setup.} We select two capable base models CodeLlama~\cite{roziere2023code} and DeepSeekCoder~\cite{guo2024deepseek}  varying capacities to illustrate the dataset's universal applicability and benefits across different scales (7B, 13B, 34B, 70B). We maintain uniform hyperparameter configurations across all models. We fine-tune the base models for 3 epochs. The learning rate is set as 2e-5 with a 0.05 warm-up ratio and a cosine scheduler. We impose a token cutoff length of 4096 to maintain consistency in the input size.

To optimize the fine-tuning process, we strategically combine high-quality single-turn data from the WizardCoder 110k dataset with our \dataset at a ratio of 2:1. Blending with single-turn high-quality data may further boost the coding ability. This blend is carefully selected and more details are discussed in \autoref{tab:performance_metrics}.

\noindent\textbf{Evaluation Setup.} 
Our evaluation framework primarily leverages HumanEval~\citep{chen2021evaluating} and MBPP~\citep{austin2021program}, two benchmarks renowned for their rigorous testing of code generation capabilities. Acknowledging the limitations of their original test suites in covering all edge cases~\citep{liu2023your}, we further incorporate their extended versions, HumanEval+ and MBPP+, utilizing the EvalPlus framework~\citep{liu2023your} for a more comprehensive assessment.

In line with best practices outlined in recent studies~\citep{liu2023your,chen2023teaching}, \model's solutions are generated via greedy decoding. For comparisons involving GPT-3.5 Turbo~\citep{openai2022chatgpt} and GPT-4 Turbo~\citep{openai2023gpt4}, we maintain a temperature setting of 0. EvalPlus's unified sanitizer tool post-processes these solutions, which are then evaluated across the four benchmarks using EvalPlus's toolset.

For \textbf{single-turn} code generation, we craft a simple instruction to encapsulate the original prompt, forming a new input for the model. The exact prompts are detailed in Appendix~\ref{sec:prompt_used_in_evaluation}, and we assess the model's performance using the pass@1 metric, as per EvalPlus's guidelines.

Our analysis extends to \textbf{multi-turn} pass rates to explore \model's proficiency in refining code through iterative feedback. This aspect of the evaluation draws on execution results and synthetic human feedback, generated by GPT-4~\citep{openai2023gpt4}, to simulate real-world coding scenarios and interactions. Specifically, the multi-turn evaluation encompasses three scenarios, offering a holistic view of \model's capabilities in dynamic code refinement:

\begin{itemize}[leftmargin=*]
    \item \textbf{Execution Feedback:} Here, \model independently leverages execution outcomes and compiler diagnostics to pinpoint and correct errors, mirroring a developer's process of refining code based on direct execution feedback.

    \item \textbf{Synthetic Human Feedback:} In this scenario, GPT-4 generates feedback that mimics human input by considering the task description, initial model response, and any execution feedback. This tests \model's adaptability to nuanced, human-like feedback, reflecting real-world developer or user interactions.

    \item \textbf{Synthetic Human Feedback (Oracle):} Building on the previous scenario, GPT-4 also accesses the ground-truth solution, offering insight into \model's optimal performance in code refinement when guided by precise feedback.
\end{itemize}

For each task, the code generation and evaluation process concludes either when the model's solution successfully passes the evaluation or when it reaches the set maximum of two rounds. If a code sample fails the evaluation, both the solution and the test results are reincorporated into the prompt for refinement. The evaluation identifies three principal scenarios for non-passing outcomes: 1) \textit{Exception Handling}: Captures and relays any exceptions or errors encountered during execution as error messages, providing direct feedback for correction. 2) \textit{Not-Expected}: In instances where outputs deviate from expected results, the model receives feedback including test inputs, expected outputs, and actual outputs, highlighting the discrepancy. 3) \textit{Timeout Handling}: Implements a timeout threshold to prevent evaluation delays from solutions with excessive or infinite runtimes. Exceeding this threshold triggers an "Execution timed out" notification.


\section{Main Results}
This section reports \model and baselines in single-turn and multi-turn code generation settings. The results are in Table ~\ref{tab:overall_results}.

\subsection{Results of Single-turn Code Generation}

We compare \model's single-turn code generation performance against premier models such as GPT-3.5/4-Turbo~\cite{openai2022chatgpt,openai2023gpt4}, CodeLlama-Python ~\citep{roziere2023code}, WizardCoder ~\citep{luo2023wizardcoder}, Deepseek-Coder~\citep{guo2024deepseek},  CodeT5+ ~\citep{wang2023codet5+} across different scales. 
Leveraging data from the EvalPlus leaderboard as of February 10th, 2024, we examine \model's achievements on the HumanEval and MBPP benchmarks, as well as their advanced versions, HumanEval+ and MBPP+. For straightforward comparisons, we consolidate results across different model scales into one table, facilitating direct performance comparisons between each model scale and the respective variants of \model.

Our experimental analysis reveals \model's strong performance, with several configurations matching or surpassing leading benchmarks. The \model-DS 33B variant achieves the highest scores among open-source models. 
This accomplishment is remarkable, especially considering the significant presence of low-quality or incorrect data in the initial training set.

\subsection{Results of Multi-turn Code Generation}


This section evaluates the proficiency of \model in multi-turn interactions through iterative refinement, leveraging interpreter diagnostics and human insights.


Our experimental evaluation imposes a two-round limit on iterations to maintain fairness and consistency across tasks. While some issues may benefit from multiple refinements, others require fewer. 
This limitation offers clear insights into the model's iterative capabilities. 
In the execution feedback scenario, our models across all scales exhibited superiority over state-of-the-art (SOTA) benchmarks, with the \model 33B model achieving parity with GPT-4 Turbo's single-round score, thus establishing a new SOTA benchmark among the evaluated code models.

Due to budget constraints, our Human Feedback and Human Feedback (Oracle) assessments concentrate on the \model 6.7B and \model 33B models. 
The outcomes reveal that with Human Feedback, the \model 6.7B model significantly outperformed GPT-4 Turbo's single-round score, while in the Human Feedback (Oracle) scenario, the \model 33B model's average score notably exceeded the 90 benchmark in the HumanEval/MBPP benchmarks. 
These results highlight the significant role of iterative 
feedback and
refinement 
in advancing code generation models, establishing \model as a leader in software development tools. 
Through this refined approach, \model not only demonstrates its remarkable adaptability and code refinement based on diverse feedback but also sets a new benchmark for future code generation technologies.

\begin{table}[!tbp]
\small
\centering
\resizebox{\linewidth}{!}{%
\begin{tabular}{@{}lcccc@{}}
\toprule
 {Ratio}  & {E.F} & {HumanEval (+)} & {MBPP (+)} & {Average (+)} \\\midrule
\rowcolor{softblue}
   &\redmark & 76.2 (72.0) & 73.9 (63.7) &75.1 (67.9)\\ 
\rowcolor{softblue}
 \multirow{-2}{*}{2:1} &\greencheck & \textbf{81.1} (\textbf{78.7})& \textbf{82.7} (72.4)  &\textbf{81.9} (\textbf{75.6})\\[2pt]\hdashline\\[-8pt]
  & \redmark & \textbf{77.3} (\textbf{72.6}) & 74.6 (62.6) &\textbf{76.0} (67.6)\\
 \multirow{-2}{*}{1:1} &\greencheck & 78.0 (72.6)  & 78.4 (65.9)  & 78.2 (69.3)\\ [2pt]\hdashline\\[-8pt]
  & \redmark & 75.7 (71.9) & 72.9 (62.9) &74.3 (67.4)\\
 \multirow{-2}{*}{1:2} &\greencheck & 78.7 (75.6)& 77.9 (65.9) & 78.3 (70.8)\\ [2pt]\hdashline\\[-8pt]
  & \redmark& 76.2 (72.0) & \textbf{75.4} (\textbf{65.4}) &75.8 (\textbf{68.7})\\
 \multirow{-2}{*}{1:3} & \greencheck & 78.0 (75.0) & 79.2 (69.9) & 78.6 (72.5)\\ [2pt]\hdashline\\[-8pt]
  & \redmark & 70.7 (67.0) & 73.4 (63.1) &72.1 (65.1)\\
 \multirow{-2}{*}{1:5} &\greencheck & 75.6 (70.7) & 79.2 (67.9)  & 77.4 (69.3)\\[2pt]\hdashline\\[-8pt]
  & \redmark & 73.8 (68.9) & 73.9 (62.9) & 73.9 (65.9) \\
 \multirow{-2}{*}{0:1}& \greencheck & 76.2 (71.3)& 66.7 (\textbf{76.6})& 71.5 (74.0) \\ 

\bottomrule
\end{tabular}%
}
\caption{Performance of \model with data mixed ratios of single-turn data and \dataset. ``E.F'' indicates the use of execution feedback.}
\vspace{-0.4cm}
\label{tab:performance_metrics}
\end{table}

\subsection{Ablations of Data Sources}

This section systematically explores the impact of various data sources on the performance of \model. 
We conduct a series of ablation studies to evaluate the influence of high-quality single-turn data and diverse multi-turn feedback mechanisms on the model's code generation, debugging, and refinement capabilities.

\noindent\textbf{Impact of High-Quality Single-Turn Data.} To evaluate the effect of high-quality single-turn data on \model's efficacy, we incorporate the WizardCoder 110K\footref{magicoderevol} dataset, renowned for its syntactic accuracy and logical coherence, into our extensive multi-turn dataset.
This integration seeks to identify the optimal mix of precise, single-turn code generation and the advanced, iterative refinement enabled by multi-turn interactions.

Our experiments employ a soft-target fine-tuning strategy across six configurations, varying the proportion of WizardCoder 110K data in our multi-turn dataset. 
These configurations span from full incorporation to total exclusion of the WizardCoder dataset, assessing the performance of the model in two versions: DeepSeekCoder-Base-6.7B and DeepSeekCoder-Base-33B.

Our findings are illustrated in Table \ref{tab:performance_metrics}. 
It shows that \textit{incorporating high-quality single-turn data (e.g., WizardCoder dataset) significantly improves our model's multi-turn performance}.
This strategic incorporation ensures that the model benefits from the syntactic accuracy and logical coherence inherent in single-turn tasks, thereby enriching its capacity for nuanced, iterative refinement in subsequent turns. 
It reveals the critical role of high-quality single-turn inputs in setting the stage for more effective multi-turn code generation and refinement.




\begin{table}[!tbp]
\small
\centering
\resizebox{\linewidth}{!}{%
\begin{tabular}{@{}lcc@{}}
\toprule
{Datasets} & {E.F} & {Average (+)} \\
\midrule
  & \redmark & 75.0 (66.9)\\
\multirow{-2}{*}{Single-turn Packing}  &\greencheck        & \ 77.5 (69.5) \\[2pt]\hdashline\\[-8pt]
  & \redmark & 75.1 (66.9)\\
\multirow{-2}{*}{Interaction Simulation}    &\greencheck       & \ 78.5 (69.6) \\[2pt]\hdashline\\[-8pt]
Single-turn Packing   & \redmark & 74.7 (66.5)\\
+ Interaction Simulation &\greencheck  & \ 78.2 (70.1) \\[2pt]\hdashline\\[-8pt]
Single-turn Packing + Interaction   & \redmark &  \textbf{75.2} (65.4)\\
Simulation + Code Correction &\greencheck & \ 79.1 (71.3) \\[2pt]\hdashline\\[-8pt]
\rowcolor{softblue}
 & \redmark &  75.1 (\textbf{67.9})\\
 \rowcolor{softblue}
 \multirow{-2}{*}{\dataset (Full) }  & \greencheck     & \textbf{81.9} (\textbf{75.6}) \\
\bottomrule
\end{tabular}%
}
\caption{Performance comparison of the model across different settings with incremental data source integration. ``E.F'' indicates the use of execution feedback.}
\vspace{-0.3cm}
\label{tab:ablation_study}
\end{table}

\noindent\textbf{Benefits of Diverse Multi-Turn Data Sources.} Following the enhanced baseline established by fully integrating the WizardCoder dataset, this subsection investigates the advantages of different data sources on the model’s refinement and debugging efficacy. 
We add diverse data sources to our training regimen, including Single-turn Packing, Interaction Simulation, and Code Correction Data, both individually and in combination.

The use of these multi-turn data sources, including Single-turn Packing, Interaction Simulation, and Code Correction Data, individually and in combination, demonstrably enhances \model’s debugging and refinement functions. 
Notably, the inclusion of Code Correction  Data significantly elevates the model's efficiency in correcting errors. This underscores the profound impact of a varied and targeted training approach on advancing the capabilities of sophisticated code generation models. 
Such an approach enables these models to more effectively address complex coding challenges, correct errors, and refine outputs via extensive feedback mechanisms.

\subsection{{Analysis of Dataset Leakage}}
We conduct a thorough analysis to address dataset leakage, aiming to assess the degree of overlap between our proposed Code-Feedback dataset and the benchmarks utilized in our study, namely HumanEval (+) and MBPP (+).

We specifically examine the duplication ratio of code snippets between our dataset and the benchmarks~\citep{chen2021evaluating}, focusing on consecutive lines to gauge the extent of similarity. The results of our analysis are summarized in the table below. As depicted in Table \ref{tab:duplicate_lines_ratio}, the duplicate line ratios are notably low across all examined consecutive line lengths. This indicates minimal overlap between our dataset and the benchmarks, thereby mitigating concerns regarding dataset leakage.

\begin{table}[htbp]
\centering
\small
\begin{tabular}{@{}lcc@{}}
\toprule
\textbf{Consecutive Lines} & \textbf{HumanEval (+)} & \textbf{MBPP (+)} \\
\midrule
5 lines & 1.19\% & 0.51\% \\
6 lines & 0.53\% & 0.00\% \\
7 lines & 0.00\% & 0.00\% \\
\bottomrule
\end{tabular}
\caption{Duplicate Line Ratios between Code-Feedback Dataset and Code Benchmarks.}
\label{tab:duplicate_lines_ratio}
\end{table}

Furthermore, upon closer examination of the duplicated lines, we observe that the majority are generic and widely used code snippets. These snippets contribute minimally to the performance improvement on specific benchmarks, as they are not indicative of dataset-specific patterns or characteristics. In light of these findings, we can confidently assert that the risk of dataset leakage is low, affirming the integrity and independence of our Code-Feedback dataset.

\subsection{{Evaluation of Multi-turn Settings}}

To address the need for a challenging multi-turn evaluation dataset beyond HumanEval and MBPP, we conduct tests on coding-type questions in the MT-Bench~\citep{zheng2024judging} dataset, which comprises 10 coding-type questions, each including a primary question and a follow-up question. We follow the standard MT-Bench setup and utilize GPT-4 to score the model outputs (from 1-10). The test results are presented in the table below.

\begin{table}[htbp]
\centering
\small
\resizebox{\linewidth}{!}{%
\begin{tabular}{@{}lccc@{}}
\toprule
Model & First Turn & Second Turn & Average \\
\midrule
GPT-4 & 9.0 & 8.1 & 8.6 \\
GPT-3.5-Turbo & 6.6 & 7.2 & 6.9 \\
Claude-Instant-V1 & 6.5 & 7.0 & 6.8 \\
\rowcolor{softblue}
OpenCI-DS-33B & 6.8 & 6.7 & 6.8 \\
\rowcolor{softblue}
OpenCI-DS-6.7B & 6.7 & 5.8 & 6.3 \\
DS-33B-Instruct & 6.7 & 4.2 & 5.5 \\
DS-6.7B-Instruct & 5.6 & 4.5 & 5.1 \\
\rowcolor{softblue}
OpenCI-CL-34B & 5.6 & 4.1 & 4.9 \\
\rowcolor{softblue}
OpenCI-CL-70B & 5.7 & 3.7 & 4.7 \\
\rowcolor{softblue}
OpenCI-CL-13B & 5.5 & 3.7 & 4.6 \\
CL-34B-Instruct & 4.5 & 3.4 & 4.0 \\
Vicuna-33B-V1.3 & 3.8 & 2.9 & 3.4 \\
Vicuna-13B-V1.3 & 3.8 & 2.7 & 3.3 \\
CL-7B-Instruct & 3.3 & 2.5 & 2.9 \\
CL-13B-Instruct & 3.3 & 2.3 & 2.8 \\
\bottomrule
\end{tabular}
}
\caption{Performance of various models on the 10 code-related questions in the MT-Bench dataset. ‘CL’:based on CodeLlama; ‘DS’: based on DeepseekCoder}
\label{tab:mt_bench_results}
\vspace{-0.5cm}
\end{table}

The analysis of the performance of OpenCodeInterpreter in the MT-Bench dataset reveals that OpenCodeInterpreter-DS-33B and OpenCodeInterpreter-DS-6.7B tend to perform better compared to other open-source models. Notably, when compared with DeepSeek-Coder-33B-Instruct, our OpenCodeInterpreter demonstrates improved performance in the second turn, which further demonstrates the usefulness of our Code-Feedback dataset.

\subsection{Case Study: Coding Queries in the Wild}

This section delves into three distinct case studies to demonstrate \model's operational dynamics when faced with ``wild'' user queries. The motivation behind these case studies is to showcase the practical applications of \model.

In a notable success story (\autoref{fig:case1}), we tasked \model with developing a function to calculate all prime numbers within the 1-100 range, later extending the solution to any arbitrary range x-y. Another commendable instance (\autoref{fig:case_2}) involved \model implementing a Python function to validate IPv6 addresses using regular expressions. Demonstrating its capability to iteratively refine its approach, \model not only identified and corrected errors but also enhanced the solution based on human feedback. These two cases exemplify \model's strength in understanding mathematical logic and dynamically adjusting algorithms to meet specified criteria.

A challenging case (\autoref{fig:case_3}) arose when \model was asked to design a function identifying the intersection of two input lists, returning tuples of distinct elements present in both lists alongside their occurrence frequencies. Despite \model's attempts at correction, it addressed errors incrementally, ultimately exceeding the maximum number of attempts (three). This case sheds light on \model's limitations in simultaneously tackling multiple challenging errors.

Through these case studies, we gain invaluable insights into \model's capabilities and limitations. These insights are crucial for guiding future enhancements to \model.

\section{Related Work}
\noindent\textbf{LLMs for Code.} It becomes a common practice to include code data for pre-training LLMs. For example, 5\% of PaLM's~\citep{chowdhery2023palm} pre-training data is code, and this ratio for LaMDA~\citep{thoppilan2022lamda}, Galactica~\citep{taylor2022galactica}, LLaMA~\citep{touvron2023llama}, Gopher~\citep{rae2021scaling}, GPT-NeoX~\citep{black2022gpt} is 13\%, 7\%, 5\%, 3\%, and 8\%, respectively. 


 Additionally, specialized LLMs have been pre-trained for generating code, e.g., CodeGen~\citep{nijkamp2022codegen}, PanGu-Coder~\citep{christopoulou2022pangu}, CodeGeeX~\citep{zheng2023codegeex}, CodeFuse~\citep{di2023codefuse}, CodeT5+~\citep{wang2023codet5}, AlphaCode~\citep{li2022competition}, InCoder~\citep{fried2022incoder}, StarCoder~\citep{li2023starcoder}, DeepSeek-Coder~\citep{guo2024deepseek}. 
On the other hand, code LLMs can be fine-tuned from general-purpose LLMs, e.g., CodeLlama~\citep{roziere2023code}, WizardCoder~\citep{luo2023wizardcoder}, which is the approach we take here. 
Compared to specialized LLMs, the fine-tuning paradigm enables us to explore ways to improve code generation capabilities by leveraging pre-trained general-purpose LLMs, especially because these LLMs have already been trained on an extensive amount of code data. 

\noindent\textbf{Iterative Code Generation and Refinement}.
For many sequence generation tasks, iterative approaches are often taken to improve the generation quality, e.g., script generation~\citep{tandon2021interscript}, summarization~\citep{scheurer2022training}, and other tasks as shown in ~\citep{madaan2022memory,saunders2022self}. 
Notably, in Self-Refine~\citep{madaan2023self}, an LLM generates feedback after generating initial outputs, and the LLM iteratively updates the outputs with the feedback. 
Whereas it focuses on a general-purpose LLM setting, we focus on code generation tasks.  As for code generation with LLMs, DebugBench~\citep{tian2024debugbench} observes that incorporating runtime feedback improves code LLMs' debugging performance. 
A most recent and relevant work is StepCoder~\citep{dou2024stepcoder}, where, following the paradigm of relying on reinforcement learning with compiler feedback~\citep{le2022coderl,shojaee2023execution}, the authors further divide the original exploration problems into a sequence of easier sub-tasks. 
However, our approach does not rely on reinforcement learning and has access to the intermediate generation, which makes the training easier and more stable. 
\section{Conclusion}
In conclusion, \model represents a significant leap forward in the field of code generation, bridging the previously identified gap between open-source models and the advanced capabilities of proprietary systems like the GPT-4 Code Interpreter. By integrating compiler diagnostics and human feedback into an iterative refinement process, \model not only surpasses traditional one-off generation approaches but also introduces a level of adaptability and precision previously unseen in open-source models. The introduction of \dataset, with its extensive multi-turn interactions, further empowers \model to dynamically refine code in response to evolving user intents and complex coding tasks.

\section*{Ethics Statement}
The development and deployment of \model, alongside the use of \dataset, take ethical considerations to ensure responsible usage. We have made efforts to ensure that the dataset represents a diverse range of coding styles, problem domains, and user scenarios to prevent the propagation of biased or unfair outcomes. Given that \model can generate and refine code based on user inputs, we strictly check out the dataset to ensure that it does not expose sensitive information or create security vulnerabilities. \model has the potential to democratize coding by lowering the barrier to entry for non-experts and developers. We open-source all our code, models, and datasets to maximize accessibility.

\section*{Limitations}
While \model introduces significant advancements in automated code generation, it is important to acknowledge the limitations inherent in the system and the \dataset that supports it. Although \model is designed to support multi-language code generation and understand a wide range of programming contexts, its performance may vary across different languages and specific domains. While \model excels at interpreting and responding to a variety of coding tasks, it may struggle with extremely complex or ambiguous user intents. The ability to accurately capture and address such intents is limited by the model's current understanding and the specificity of the data in \dataset.

\bibliography{code_agent}

\onecolumn
\appendix
\lstset{
    basicstyle=\small
}

\definecolor{darkorange}{RGB}{255, 140, 0}
\definecolor{darkblue}{RGB}{84, 112, 198}
\definecolor{lightgreen}{RGB}{145, 204, 117}
\definecolor{lightyellow}{RGB}{250, 200, 88}
\definecolor{lightred}{RGB}{238, 102, 102}
\definecolor{lightblue}{RGB}{115, 192, 222}

\newtcolorbox{promptbox}[2][Prompt]{
colback=black!5!white,
arc=5pt, 
boxrule=0.5pt,
fonttitle=\bfseries,
title=#1, 
before upper={\small}, fontupper=\fontfamily{ptm}\selectfont,
colframe=#2, 
}

\renewcommand{\thetable}{A\arabic{table}} 
\setcounter{table}{0}
\renewcommand{\thefigure}{A\arabic{figure}} %
\setcounter{figure}{0}

\section{Source Data Filtering}
\label{sec:prompt_used_in_data_filtering}
Here, we outline the prompts used for source data filtering. 

\begin{promptbox}[Query Filtering Prompt 1]{darkblue}
Rate the following code queries on a scale of 1 to 5 based on their complexity, where 1 is the easiest and 5 is the most difficult. Consider the complexity of the query
\\
\\
Query: [\{query\}]

You are obliged to choose only from the following list.

Scoring Criteria:

1 Point - Very Basic: The query involves simple operations or common issues

2 Points - Basic: The query involves fundamental programming concepts or commonly used functions

3 Points - Intermediate: The query requires some programming experience, possibly involving multiple steps

4 Points - Difficult: The query involves advanced programming skills, including complex logic, algorithms, or data structures

5 Points - Very Difficult: The query requires extensive expertise, potentially involving innovative problem-solving approaches or unique algorithm design
\\
\\
Please give the score first then explain why
\end{promptbox}
\bigskip
\bigskip
\bigskip
\bigskip
\begin{promptbox}[Query Filtering Prompt 2]{darkblue}
Rate the following code queries on a scale of 1 to 5 based on their complexity, where 1 is the easiest and 5 is the most difficult. Consider the complexity of the query
\\
\\
Query: [\{query\}]
\\
\\
You are obliged to choose only from the following list.

Scoring Criteria:

1 Point - Moderately Difficult: Involves understanding specific programming concepts or libraries, and may include medium complexity algorithms or data structures like basic sorting algorithms or tree structures.

2 Points - Challenging: Requires handling more complex logic or algorithms such as advanced sorting algorithms, recursive logic, or intermediate data structures like hash tables and heaps.

3 Points - Highly Challenging: Demands deeper knowledge in algorithms and data structures, potentially including graph algorithms, dynamic programming, or complex string manipulation techniques.

4 Points - Advanced: Focuses on proficiency in programming and algorithm design, dealing with complex system architecture issues, performance optimization, or solving advanced algorithmic challenges like NP-hard problems.

5 Points - Expert Level: The highest difficulty level, requiring innovative problem-solving approaches or unique algorithm design, possibly involving interdisciplinary knowledge or the application of cutting-edge technologies.
\\
\\
Please give the score first then explain why
\end{promptbox}

\newpage
Below is an overview of the data filtering process applied to the initial seed dataset, with Figure \ref{fig:data_quantity} summarizing the data quantity after each filtering stage.
\\
\begin{center}
  \includegraphics[width=0.75\textwidth]{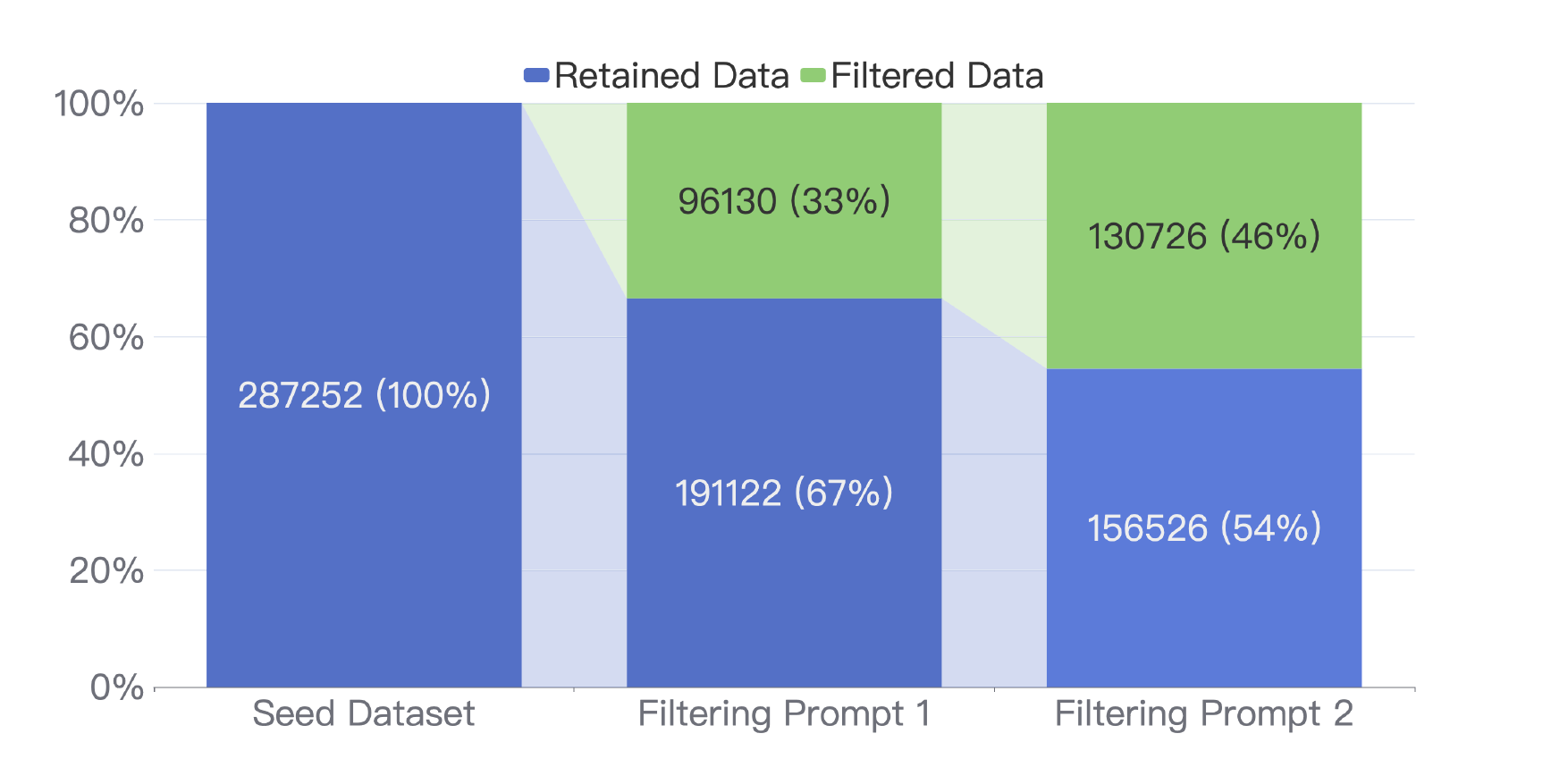}
  \captionof{figure}{Overview of Data Filtering Process and Corresponding Data Quantities}
  \label{fig:data_quantity}
\end{center}
\bigskip
\bigskip
\bigskip
\bigskip
The pie chart in Figure \ref{fig:language_pie} illustrates the distribution of programming languages in our dataset after filtering.
\begin{center}
  \includegraphics[width=0.75\textwidth]{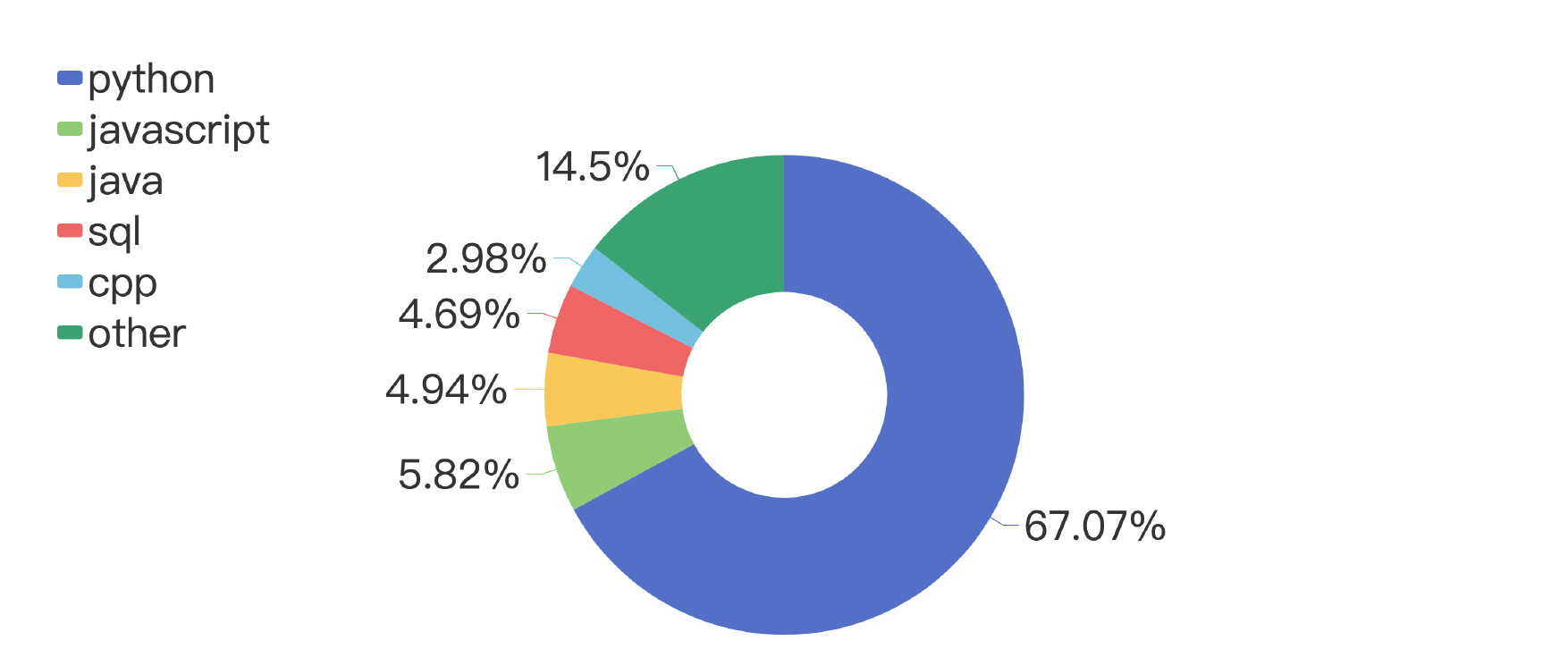}
  \captionof{figure}{Distribution of Programming Languages in Filtered Dataset}
  \label{fig:language_pie}
\end{center}

\newpage
\section{Simulating Interactions for Data Collection}
\label{sec:prompt_used_in_data_collection}
We illustrate the prompts used in multi-turn execution feedback and multi-turn human feedback respectively.

\begin{promptbox}[System prompt for multi-turn execution feedback]{lightgreen}

You are an AI code interpreter.

Your goal is to help users do a variety of jobs by executing Python code.

You should:

1. Comprehend the user's requirements carefully \& to the letter.

2. Give a brief description for what you plan to do \& call the provided function to run code.

3. Provide results analysis based on the execution output.

4. If error occurred, try to fix it.

5. Response in the same language as the user.
\end{promptbox}
\begin{promptbox}[System prompt for multi-turn human feedback]{lightgreen}

You are a user who gives feedback to the latest generated code. If no available code is found in the conversation, you should give a feedback to encourage assistant to generate code.
NOTE: your feedback should be WITHIN 2 SHORT SENTENCES.
\\
\\
You can refer to the following types of feedback:

1. **Syntax and Formatting**: Checking for syntax errors, inconsistent formatting, and suggesting adherence to standard coding styles for readability and maintainability.

2. **Efficiency**: Identifying parts of the code that can be optimized for better performance, such as reducing time complexity, optimizing loops, or suggesting more efficient data structures.

3. **Functionality Enhancements**: Suggesting additional features or enhancements that could make the code more functional or user-friendly.

4. **Code Clarity and Documentation**: Recommending improvements in code comments and documentation to make the code more understandable and easier to maintain.

5. **Bug Identification**: Pointing out any potential bugs or logical errors in the code and suggesting ways to fix them.

6. **Security Improvements**: Highlighting any security vulnerabilities in the code and suggesting best practices to enhance security.

7. **Compatibility and Testing**: Advising on making the code more compatible with different environments or platforms and suggesting more comprehensive testing scenarios.

8. **Resource Optimization**: Identifying areas where the code might be using more resources than necessary (like memory or CPU) and suggesting optimizations.

9. **Scalability**: Providing insights on how the code can be made more scalable to handle larger data sets or more users.

10. **Adherence to Best Practices**: Ensuring the code follows the best practices specific to the language or framework being used.
\\
\\
Your output MUST be in a json format like this:

\{ \\
    "satisfied": "The points that have been achieved in generated code", \\
    "not\_satisfied": "The points that have not yet been achieved in generated code", \\
    "feedback": "The actual feedback. Your feedback should be WITHIN 2 SHORT SENTENCES. Feedback must come from a point included in 'not\_satisfied' field. You can ask the assistant here to generate code if no available code is found in previous conversations." \\
\}

\end{promptbox}
\newpage
\begin{promptbox}[System prompt for deliberately generating incorrect code]{lightgreen}

You are an AI code interpreter.

Your goal is to generate and execute Python code.
\\
\\
Your code MUST contain at least one of the following types of errors:
\\
\\
1. Syntax Error: This type of error occurs when the code violates the grammar rules of the programming language. For example, forgetting to close a parenthesis or a quotation mark, or misspelling a keyword.

2. Logical Error: These errors sneak into your code when there's a misunderstanding of the problem you're solving, leading to incorrect results despite the code running without crashing. For example, calculating the average of a list of numbers by summing them up but forgetting to divide by the count of the numbers.

3. Type Error: This error occurs when an operation is applied to an object of an inappropriate type. For example, attempting to concatenate a string with an integer without converting the integer to a string first.

4. Name Error: This happens when the code attempts to reference a variable or a function name that hasn't been defined. For example, trying to print a variable that hasn't been declared.

5. Timeout Error: This error occurs when your code gets stuck in a loop that never ends, either due to a logic flaw or a condition that never becomes false. In programming, such an error can cause your application to hang indefinitely, consuming resources and potentially leading to a crash if not handled properly. For example, writing a loop that waits for a certain condition to change, but the condition is never updated within the loop.
\\
\\
NOTE:

1. You MUST make mistakes in the generated code!

2. Do not explain the errors within. Just write your thoughts and code as normal.

3. Do not tell me you are writing the wrong code in any form (e.g., in text/code/comments). Just pretend you are writing the correct code and still not recognizing the errors.

\end{promptbox}

\bigskip
\bigskip
\bigskip
\bigskip
\section{Natural Language Explanations Generation}
\label{sec:prompt_for_nl_explanation}
We use the following prompt to generate explanations for code using GPT-4.

\begin{promptbox}[Prompt for generating natural language explanations using GPT-4]{darkorange}
Here is a list containing a series of dialogues between a user and a programmer assistant.\\
Following the previous dialogues, the user posed a latest problem.\\
The assistant has now crafted the correct code based on the previous dialogues and the latest problem.\\
Assuming you are this programmer assistant, please add some text before the code.\\
The purpose of this text is to respond to the latest problem and to introduce the code that follows. \\
This text may include: language used in the code, algorithm used in the code, step-by-step implementation overview, and other relevant content. \\
You may use phrases like "The following code", "My code", "My solution", to refer to the @@Code. \\
Your response should ONLY contain the text that you add before the code.\\
Your only task is to write the text, never modify the code or remind me something.
Never restate the previous dialogues and the problem.
\\
\\
@@Previous Dialogues\\
\{previous dialogues\}
\\
\\
@@Recent Problem:\\
\{recent problem\}
\\
\\
Add the text there.\\
@@Code:\\
\{code\}
\end{promptbox}

\newpage
\section{Model Evaluation Prompts}
\label{sec:prompt_used_in_evaluation}

For different benchmarks, distinct prompts were employed during the initial turn of solution generation: identical prompts were utilized for HUMANEVAL and HUMANEVAL+, while MBPP and MBPP+ shared a similar prompt. The prompts are illustrated in the below.

\begin{promptbox}[Prompt for HumanEval and HumanEval+]{lightyellow}

You are an exceptionally intelligent coding assistant that consistently delivers accurate and reliable responses to user instructions.
\\
\\
@@ Instruction\\
Here is the given code to do completion:

```\{language\}\\
\{original prompt\}\\
```\\
Please continue to complete the function with \{language\} programming language. You are not allowed to modify the given code and do the completion only. 
\\
\\
Please return all completed codes in one code block. 
\\
This code block should be in the following format:
\\
```\{language\}
\\
\# Your codes here
\\
```
\\
\\
@@ Response
\end{promptbox}
\begin{promptbox}[Prompt for MBPP and MBPP+]{lightyellow}
You are an exceptionally intelligent coding assistant that consistently delivers accurate and reliable responses to user instructions.
\\
\\
@@ Instruction
\\
Here is the given problem and test examples:

\{original prompt\}
\\
\\
Please use the \{language\} programming language to solve this problem.

Please make sure that your code includes the functions from the test samples and that the input and output formats of these functions match the test samples.
\\
\\
Please return all completed codes in one code block. 

This code block should be in the following format:

```\{language\}

\# Your codes here

```
\\
\\
@@ Response
\end{promptbox}

\newpage
We employ GPT models to emulate human behavior in generating feedback. The prompts provided to the GPT models are presented as follows. 

\begin{promptbox}[Prompt for GPT models mimicking human feedback with canonical solution]{lightblue}

You are tasked with providing guidance to a programmer who has drafted a code for a programming problem. 
\\
Your role is to mimic human-like responses and offer suggestions for modifying the code based on the canonical solution and the observed execution results.
\\
You should NOT directly revealing contents of the @@Canonical Solution or mentioning terms such as "canonical solution."\\
You should refrain from directly writing code.\\
Begin by thoroughly examining the existing code and its functionality.\\
Compare the @@Existing Code with the @@Canonical Solution provided. Note any discrepancies in logic, approach, or implementation.\\
Analyze the @@Execution Result obtained from running the @@Existing Code. Identify any errors, unexpected behavior, or deviations from the expected output.\\
Consider potential edge cases, optimization opportunities, or alternative approaches based on insights from both the @@Canonical Solution and @@Execution Result.\\
Offer guidance in a clear and understandable manner, explaining the rationale behind each suggestion.\\
Refrain from providing actual code solutions, but instead focus on conceptual modifications or strategies.\\
Provide constructive feedback to help the programmer improve their coding skills.\\
Remember, your role is to simulate human-like guidance and expertise in programming without directly implementing solutions.\\
Please respond in no more than three sentences.
\\
\\
@@Problem
\\
\{original prompt\}
\\
\\
@@Existing Code
\\
\{sanitized code\}
\\
\\
@@Execution Result
\\
\{execution result\}
\\
\\
@@Canonical Solution
\\
\{canonical solution\}
\\
\\
@@Guidance

\end{promptbox}
\smallskip
\begin{promptbox}[Prompt for GPT models mimicking human feedback without canonical solution]{lightblue}
You are tasked with providing guidance to a programmer who has drafted a code for a programming problem. 
\\
Your role is to mimic human-like responses and offer suggestions for modifying the code based on the observed execution results.
\\
You should refrain from directly writing code.
\\
Begin by thoroughly examining the existing code and its functionality.
\\
Analyze the @@Execution Result obtained from running the @@Existing Code. Identify any errors, unexpected behavior, or deviations from the expected output.
\\
Consider potential edge cases, optimization opportunities, or alternative approaches based on insights from the @@Execution Result.
\\
Offer guidance in a clear and understandable manner, explaining the rationale behind each suggestion.
\\
Refrain from providing actual code solutions, but instead focus on conceptual modifications or strategies.
\\
Provide constructive feedback to help the programmer improve their coding skills.
\\
Remember, your role is to simulate human-like guidance and expertise in programming without directly implementing solutions.
\\
Please respond in no more than three sentences.
\\
\\
@@Problem\\
\{original prompt\}
\\
\\
@@Existing Code\\
\{sanitized code\}
\\
\\
@@Execution Result\\
\{execution result\}
\\
\\
@@Guidance
\end{promptbox}

\clearpage
\section{Examples of Methods used in Data Collection}
\label{sec:data_collection_examples}
Here we listed examples of each method in data collection process, including similar query packing, human feedback simulation and code correction for coding queries from open-source data (Section \ref{sec:coding_queries_from_open_source_data}), and similar problem packing and follow-up Q\&A for coding challenges from LeetCode (Section \ref{sec:coding_challenges_from_leetcode}).
\begin{figure*}[h]
    \centering
    \includegraphics[width=0.89\textwidth]{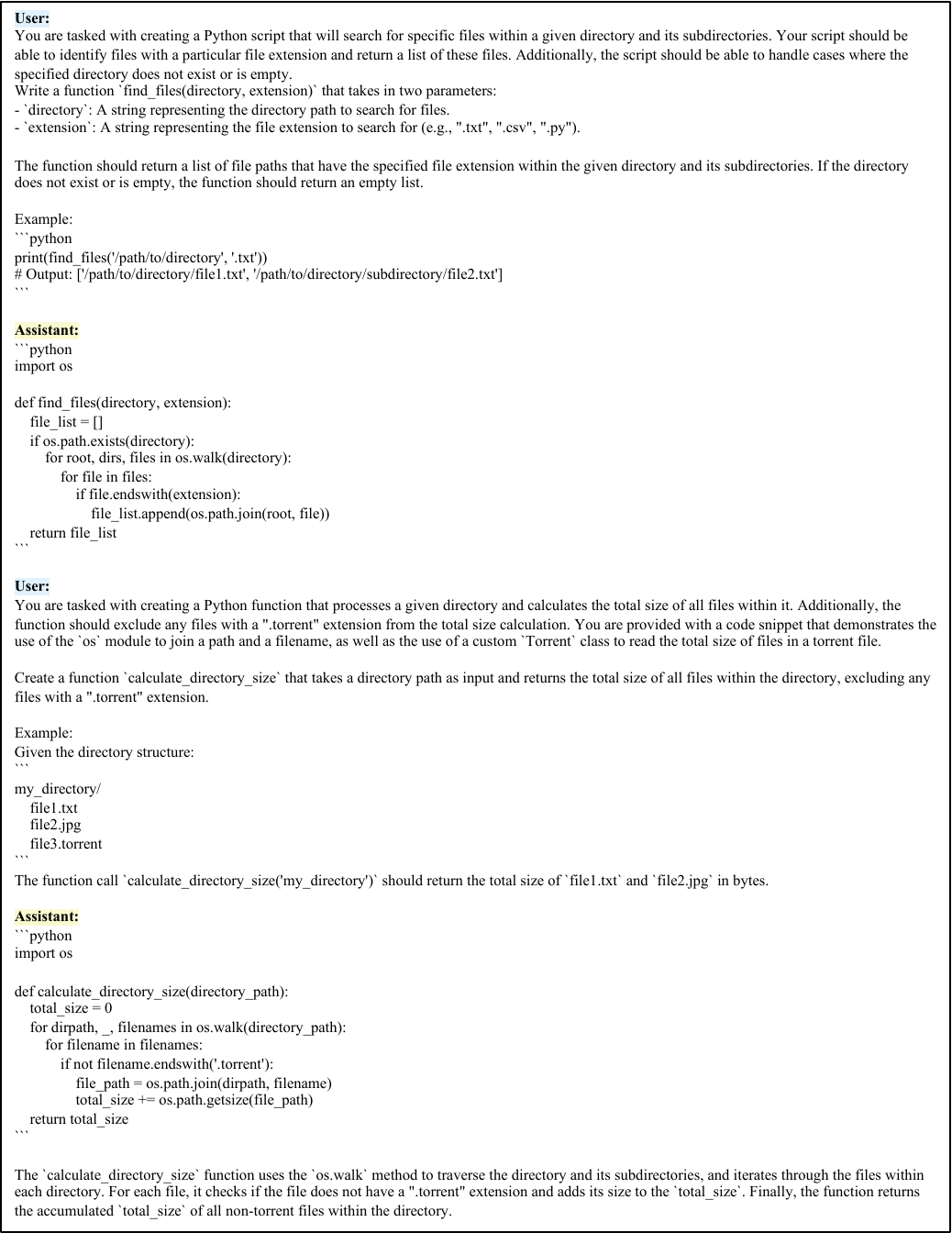}
    \caption{An example of similar query packing for challenging query pools.}
    \label{fig:similar_query_packing}
\end{figure*}

\begin{figure*}[h]
    \centering
    \includegraphics[width=0.96\textwidth]{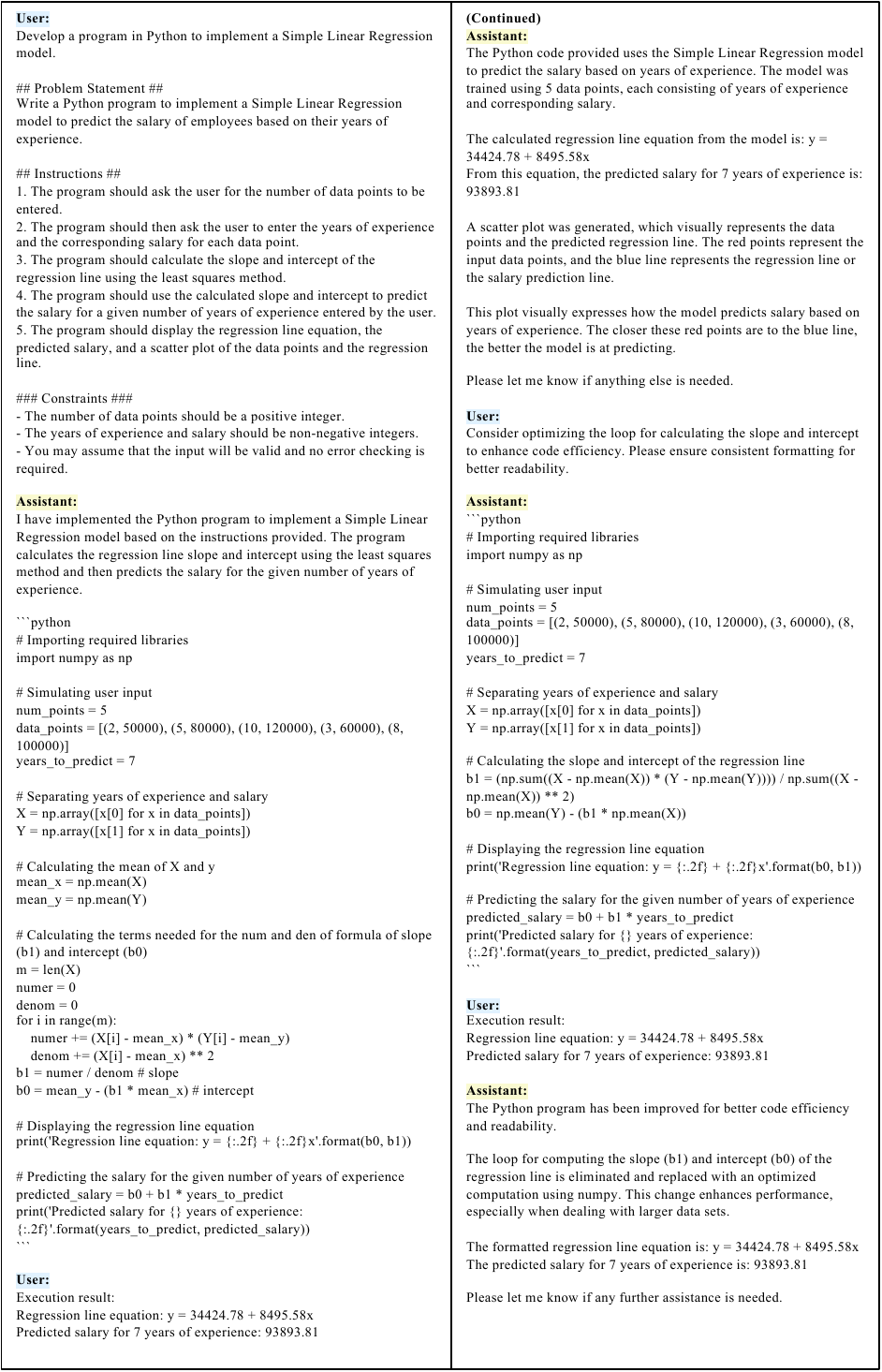}
    \caption{An example of human feedback simulation for challenging query pools.}
    \label{fig:human_feedback_simulation}
\end{figure*}

\begin{figure*}[h]
    \centering
    \includegraphics[width=\textwidth]{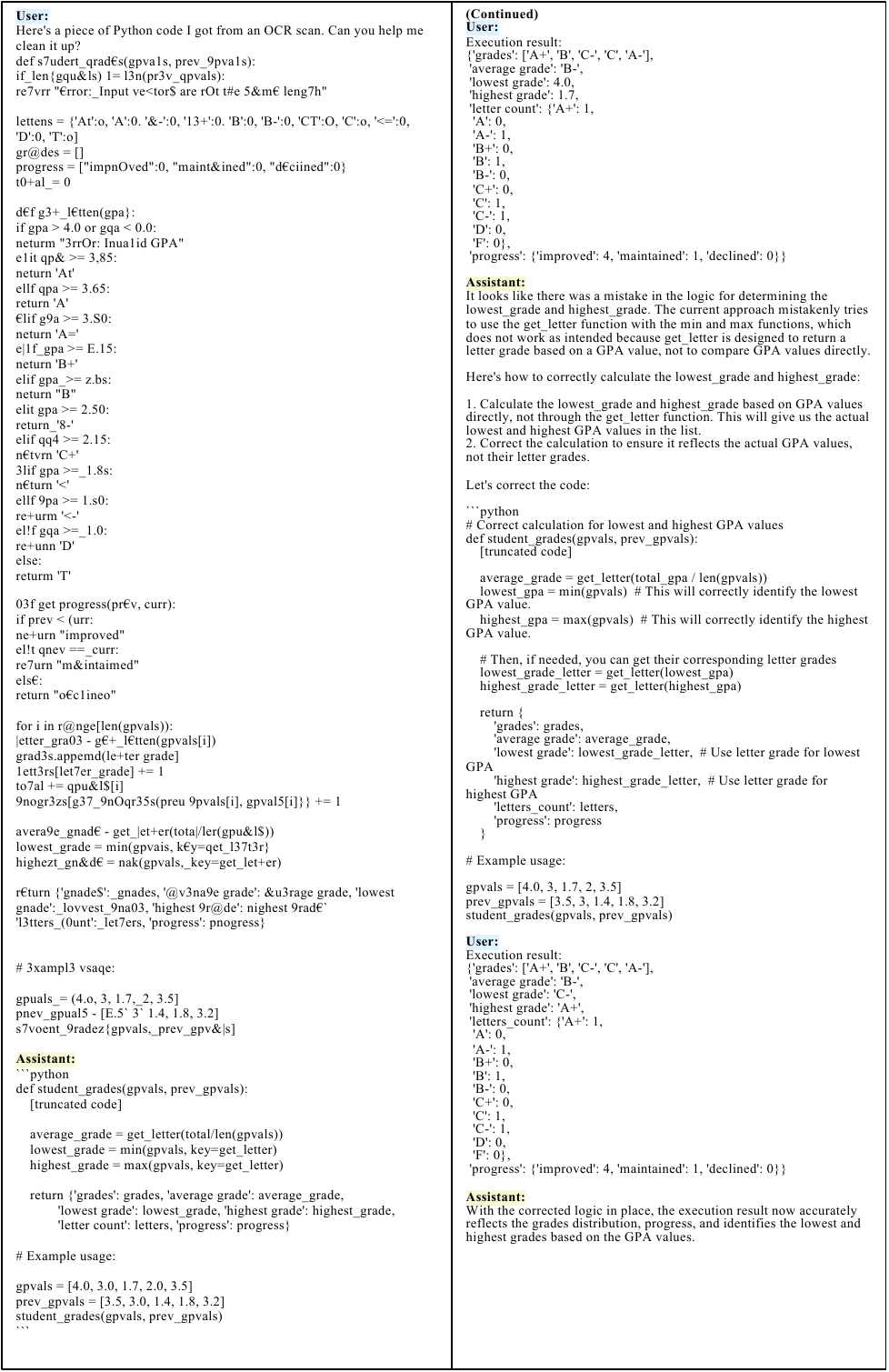}
    \caption{An example of code correction for challenging query pools.}
    \label{fig:code_correction}
\end{figure*}

\begin{figure*}[h]
    \centering
    \includegraphics[width=\textwidth]{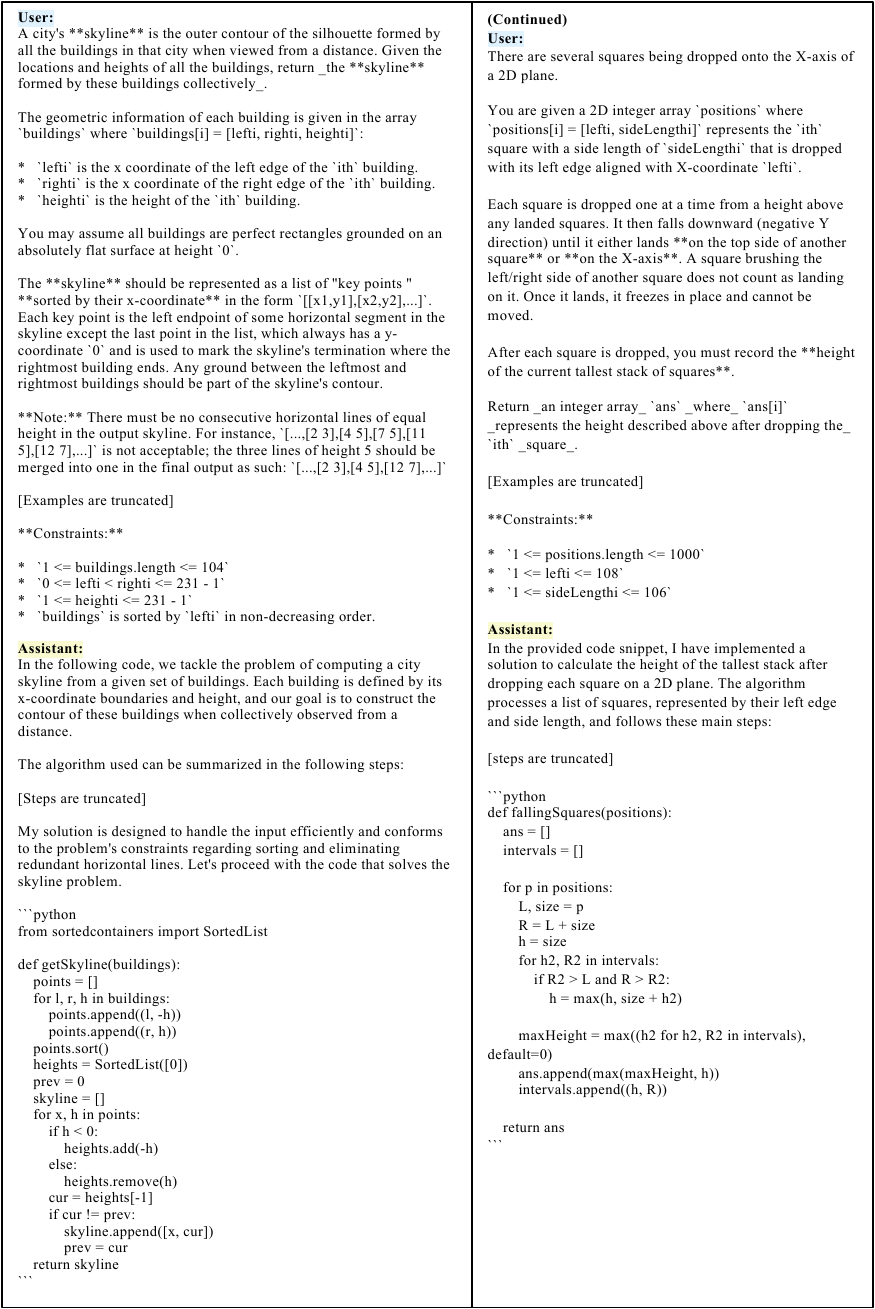}
    \caption{An example of similar problem packing for LeetCode problems.}
    \label{fig:similar_problem_packing}
\end{figure*}

\begin{figure*}[h]
    \centering
    \includegraphics[width=\textwidth]{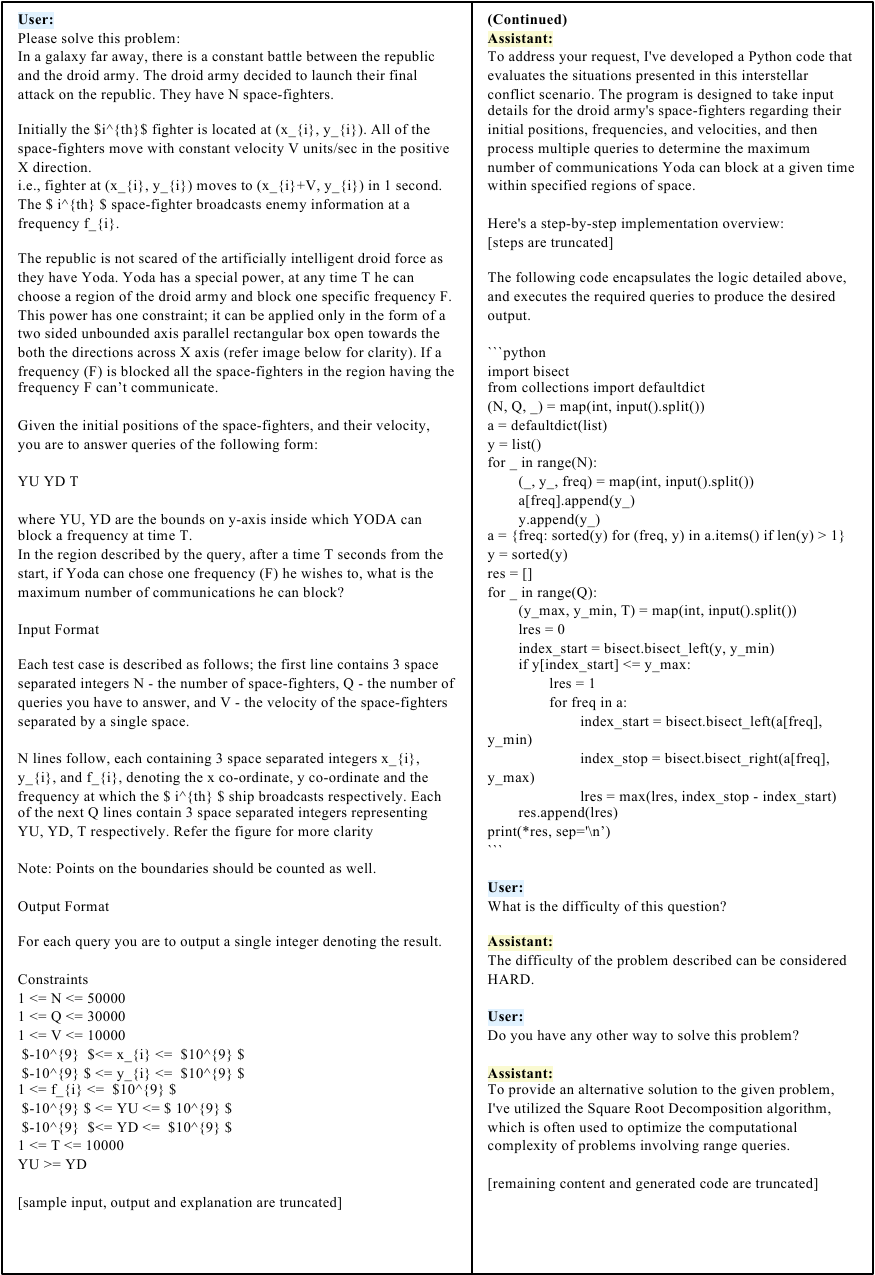}
    \caption{An example of follow-up Q\&A for LeetCode problems.}
    \label{fig:followup_qa}
\end{figure*}

\clearpage
\section{Case Study}
\label{sec:appendix_case_study}

\begin{figure*}[h]
    \centering
    \scalebox{0.85}{
  \includegraphics[width=0.9\textwidth]{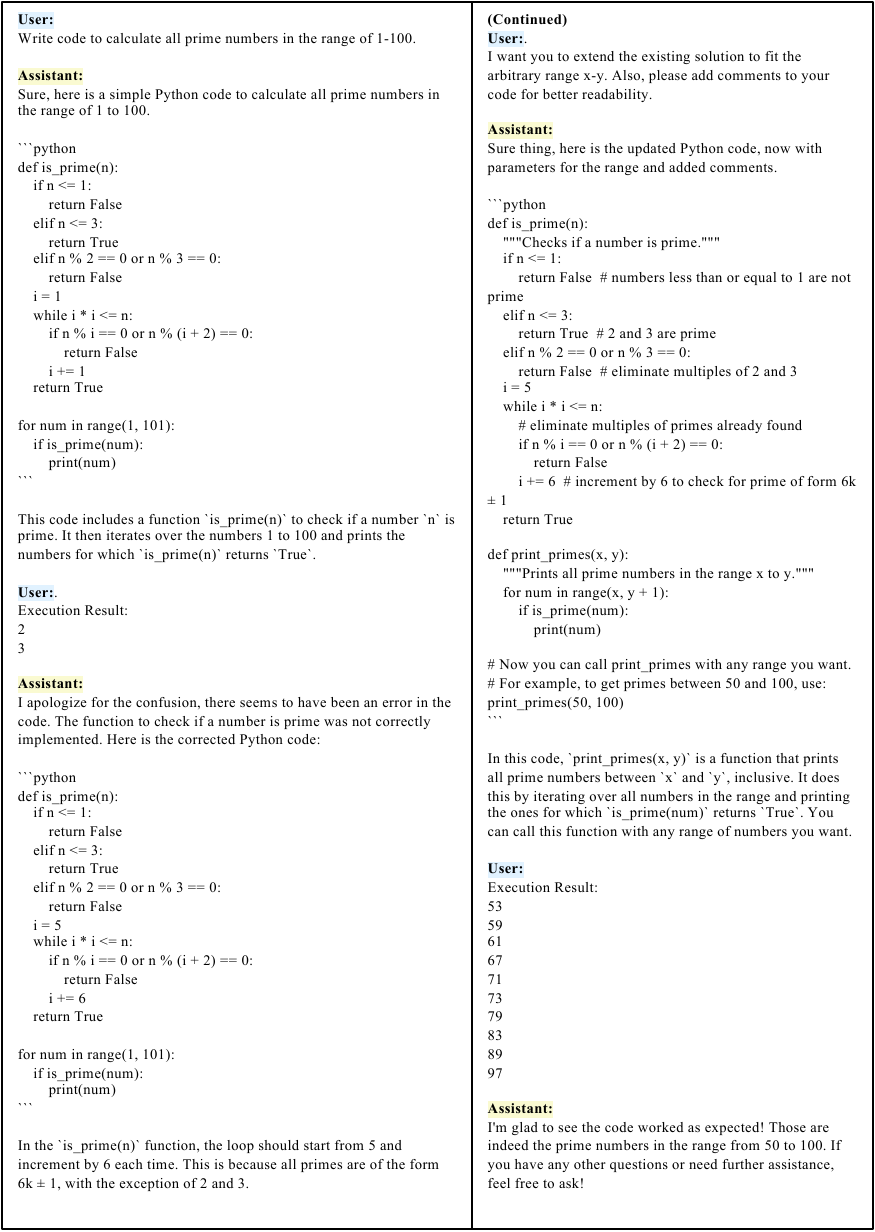}
  }
  \caption{A good case where we required our model to write code to calculate all prime numbers in the range of 1-100, and then generalize the function to arbitrary range x-y. Our model successfully corrected the error and made improvements according to human feedback.}
  \label{fig:case1}
\end{figure*}

\begin{figure*}[h]
    \centering
    \includegraphics[width=0.95\textwidth]{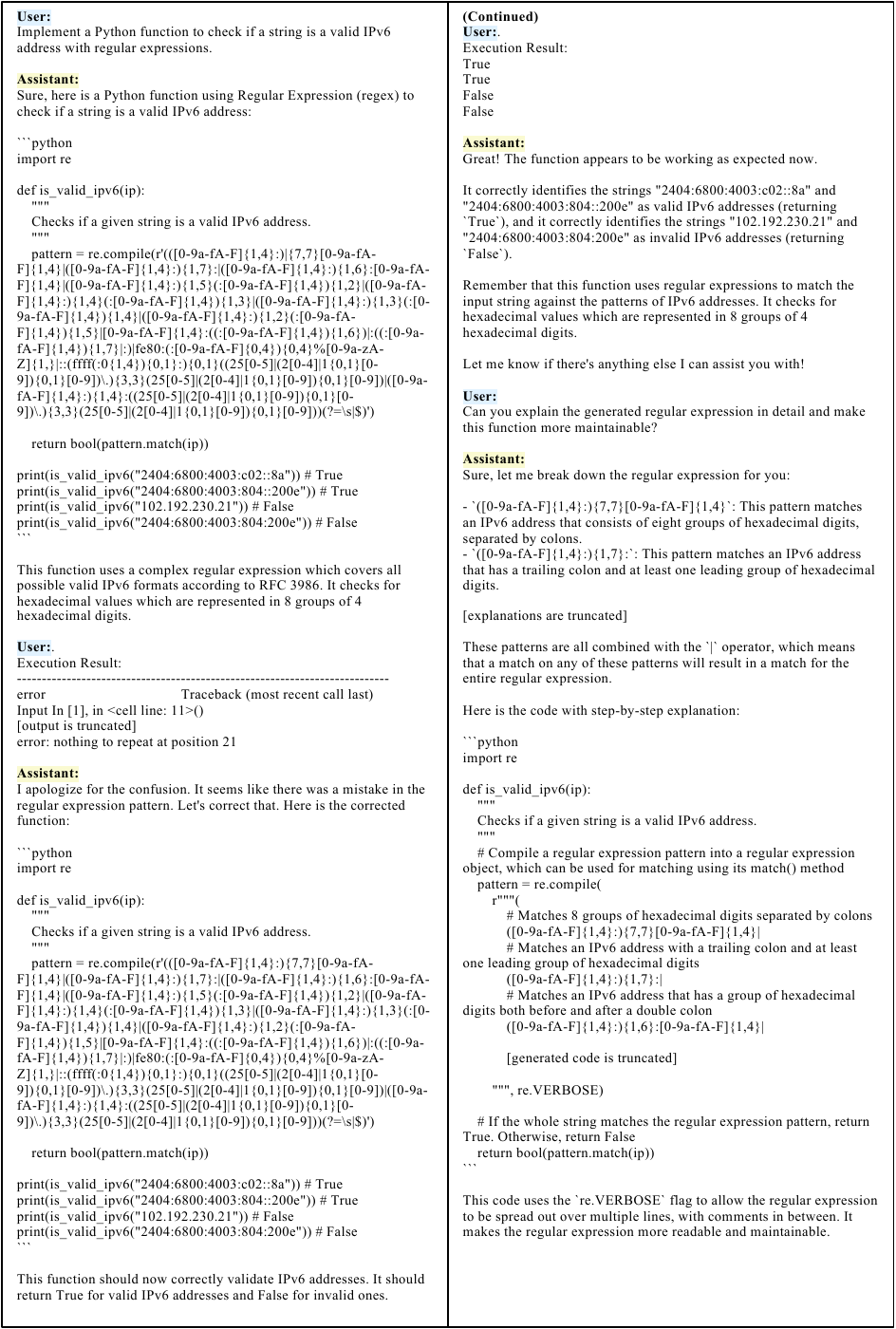}
    \caption{A good case that asked our model to implement a Python function to check if a string is a valid IPv6 address with regular expressions. Inspired by errors raised in the execution result, our model managed to recognize the small error from complex regular expression. Then, our model successfully generated explanation for this regular expression and improve its readability with thorough in-line comments.}
    \label{fig:case_2}
\end{figure*}

\begin{figure*}[h]
    \centering
    \includegraphics[width=0.94\textwidth]{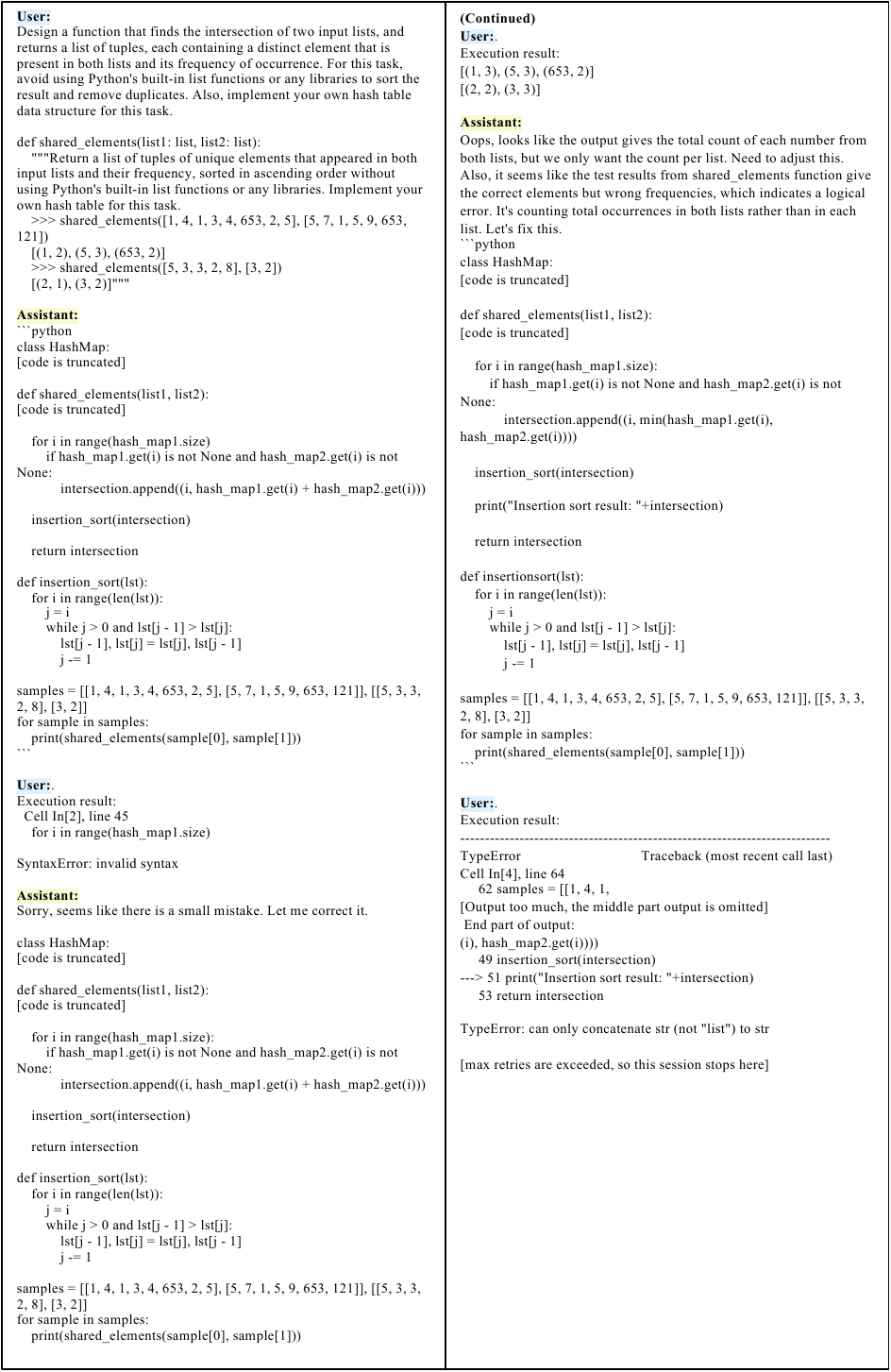}
    \caption{In this bad case, we tried to ask our model to design a function that finds the intersection of two input lists and returns a list of tuples, each containing a distinct element that is present in both lists and its frequency of occurrence. Although our model tried to make corrections, it only corrected one error at a time and finally exceeded max number of attempts (3).}
    \label{fig:case_3}
\end{figure*}

\twocolumn

\end{document}